\documentclass[12pt]{article}

\usepackage[dvips]{graphicx,color}
\newcommand{\be}{\begin{equation}}
\newcommand{\ee}{\end{equation}}
\newcommand{\ba}{\begin{eqnarray}}
\newcommand{\ea}{\end{eqnarray}}
\newcommand{\baa}{\begin{eqnarray*}}
\newcommand{\eaa}{\end{eqnarray*}}
\newcommand{\lab}[1]{\label{#1}}

\newcommand{\dis}{\displaystyle}
\newcommand{\biq}{\mbox{\boldmath $q$}}
\newcommand{\bip}{\mbox{\boldmath $p$}}
\newcommand{\bif}{\mbox{\boldmath $f$}}
\newcommand{\biL}{\mbox{\boldmath $\Lambda$}}
\newcommand{\bil}{\mbox{\boldmath $\lambda$}}

\newcommand{\bs}[1]{\mbox{\boldmath $#1$}}

\begin{document}

{\pagestyle{empty}

{\renewcommand{\thefootnote}{\fnsymbol{footnote}}
\centerline{\large \bf Generalized-Ensemble Algorithms for 
Protein Folding Simulations}
}
\vskip 1.0cm

\centerline{Yuji Sugita,$^{a,}$\footnote{\ \ e-mail: sugita@riken.jp} 
Ayori Mitsutake,$^{b,}$\footnote{\ \ e-mail: ayori@rk.phys.keio.ac.jp}
and Yuko Okamoto$^{c,}$\footnote{\ \ e-mail: okamoto@phys.nagoya-u.ac.jp} }
\vskip 1.0cm
\centerline{$^a${\it Theoretical Biochemistry Laboratory}}
\centerline{{\it Discovery Research Institute, RIKEN}}
\centerline{{\it Wako-shi, Saitama 351-0198, Japan}}
\centerline{$^b${\it Department of Physics, Keio University}}
\centerline{{\it Yokohama, Kanagawa 223-8522, Japan}}
\centerline{$^c${\it Department of Physics, Nagoya University}}
\centerline{{\it Nagoya, Aichi 464-8602, Japan}}

\vskip 1.0cm

\leftline{\hspace*{2cm} To appear in {\it Lecture Notes in Physics}}
\leftline{\hspace*{2cm} {\it Rugged Free Energy Landscapes: Common Computational Approaches}}
\leftline{\hspace*{2cm} {\it to Spin Glasses, Structural Glasses and Biological Macromolecules,}}
\leftline{\hspace*{2cm} W. Janke (ed.) (Springer-Verlag)}

%
%
\vskip 1.0cm

\centerline{\bf ABSTRACT}
\vskip 0.3cm

Conventional simulations 
of complex systems 
in the canonical ensemble suffer from the quasi-ergodicity
problem.
A simulation in generalized ensemble 
overcomes this difficulty by
performing a random walk
in potential energy space and other parameter space.
From only one simulation run, one can
obtain canonical-ensemble averages of physical quantities as
functions of temperature
by the single-histogram and/or multiple-histogram reweighting techniques.
In this article we review the generalized-ensemble algorithms.
Three well-known methods, namely, 
multicanonical algorithm, simulated tempering, and 
replica-exchange method, are described first.
Both Monte Carlo and molecular dynamics versions of the 
algorithms are given.
We then present 
further extensions of the above three methods.
}
\section{INTRODUCTION}


Canonical fixed-temperature 
simulations of complex systems such as spin glasses\index{spin glass} and 
biopolymers\index{biopolymer} are greatly hampered by the multiple-minima
problem, or the quasi-ergodicity problem.
Because simulations at low temperatures tend to get
trapped in a few of a huge number of local-minimum-energy states
which are separated by high energy barriers,
it is very difficult to obtain accurate canonical distributions
at low temperatures by conventional Monte Carlo\index{Monte Carlo} (MC)\index{MC} and
molecular dynamics\index{molecular dynamics} (MD)\index{MD} methods.
One way to overcome this multiple-minima
problem is to perform a simulation in a {\it generalized ensemble}\index{generalized ensemble} 
where each state is weighted by an artificial,
non-Boltzmann probability
weight factor so that
a random walk\index{random walk} in potential energy space may be realized
(for reviews see, e.g., Refs.~\cite{RevHO1}--\cite{RevIOO}).
The random walk allows the simulation to escape from any
energy barrier and to sample much wider configurational
space than by conventional methods.
Monitoring the energy in a single simulation run, one can
obtain not only
the global-minimum-energy state but also canonical ensemble
averages as functions of temperature by the single-histogram \cite{FS1}
and/or multiple-histogram \cite{FS2,WHAM} reweighting techniques\index{reweighting
techniques}
(an extension of the multiple-histogram method is also referred to as
{\it weighted histogram analysis method} (WHAM)\index{WHAM} \cite{WHAM}).
Besides generalized-ensemble algorithms,\index{generalized-ensemble algorithm} which are usually
based on local updates, methods based on non-local updates
such as cluster algorithms and their generalizations have
also been widely used \cite{SW87}--\cite{ELM93}.
In this article, we focus our discussion on generalized-ensemble
algorithms.

One of the most well-known generalized-ensemble methods is perhaps
{\it multicanonical algorithm}\index{multicanonical algorithm} (MUCA\index{MUCA}) \cite{MUCA1,MUCA2}
(for reviews see, e.g., Refs.~\cite{MUCArev,RevJanke}).
(The method is also referred to as {\it entropic sampling}
\cite{Lee} and {\it adaptive umbrella sampling} \cite{MZ} 
{\it of the potential energy} \cite{BK}.
MUCA can also be considered as a sophisticated, ideal realization of
a class of algorithms called {\it umbrella sampling}\index{umbrella sampling}
\cite{US}.  Also closely related methods are {\it transition
matrix methods} reviewed in Refs.~\cite{WS02,Be02} and
{\it random walk algorithm} \cite{Landau1,Landau2}, which is also 
referred to as {\it density of states Monte Carlo} \cite{dePablo}.
See also Ref.~\cite{THT04}.)
MUCA and its generalizations have been applied to
spin systems
(see, e.g., Refs.~\cite{MUCA3}--\cite{BBJ00}).
MUCA was also introduced to the molecular simulation field
\cite{HO}.
Since then MUCA and its generalizations have been extensively
used in many applications in
protein\index{protein} and related systems \cite{HO94}--\cite{OO03}.
Molecular dynamics version of MUCA
has also been developed \cite{HOE96,NNK,BK} (see also
Refs.~\cite{Muna,HOE96} for Langevin dynamics version).
MUCA has been extended so that flat distributions in
other parameters instead of potential energy may be 
obtained \cite{BeHaNe93,JK95,KPV,BK2,ICK,BNO03}.
Moreover, multidimensional (or multicomponent) extensions of MUCA
can be found in Refs.~\cite{KPV,BK2,HNSKN,OO03}.
  
While a simulation in multicanonical ensemble performs a free
1D random walk in potential energy space, that in
{\it simulated tempering}\index{simulated tempering} (ST)\index{ST} \cite{ST1,ST2} 
(the method is also referred to as the
{\it method of expanded ensemble}\index{expanded ensemble} \cite{ST1})
performs a free random walk in temperature space
(for a review, see, e.g., Ref.~\cite{STrev}).
This random walk, in turn,
induces a random walk in potential energy space and
allows the simulation to escape from
states of energy local minima.
ST has also been applied to protein folding\index{protein folding}
problem \cite{IRB1,HO96a,HO96b,IRB2}.


The generalized-ensemble algorithms are
powerful, but in the above two methods the probability
weight factors are not {\it a priori} known and have to be
determined by iterations of short trial simulations.
This process can be non-trivial and very tedius for
complex systems with many degreees of freedom.
Therefore, there have been attempts to accelerate
the convergence of the iterative process for MUCA
weight factor determination
\cite{MUCA3,KPV,SmBr,H97c,MUCAW,BK}
(see also Refs.~\cite{MUCArev,Janke03}).
  

In the {\it replica-exchange method}\index{replica-exchange method} (REM)\index{REM}
\cite{RE1}--\cite{RE2}, the difficulty of weight factor
determination is greatly alleviated.  (A closely related method 
was independently developed in Ref.~\cite{RE3}.
Similar methods in which the same equations are used but
emphasis is laid on optimizations have
been developed \cite{KT,JWK}. 
REM is also referred to as 
{\it multiple Markov chain method}\index{multiple Markov chain method} \cite{RE4}
and {\it parallel tempering}\index{parallel tempering} \cite{STrev}.  Details
of literature
about REM and related algorithms can be found in
recent reviews \cite{IBArev,RevMSO}.)
In this method, a number of
non-interacting copies (or replicas) of the original system
at different temperatures are
simulated independently and
simultaneously by the conventional MC or MD method. Every few steps,
pairs of replicas are exchanged with a specified transition
probability.
The weight factor is just the
product of Boltzmann factors, and so it is essentially known.

REM has already been used in many applications in protein 
systems \cite{H97,SO,IRB2}\cite{WD}--\cite{LMMO03}. 
Other molecular simulation fields have also been studied by
this method in various ensembles \cite{FD}--\cite{OKOM}.
Moreover, REM was applied to cluster studies in quantum
chemistry field \cite{ISNO}.
The details of molecular dynamics algorithm have been worked out
for REM in Ref.~\cite{SO} (see also Refs.~\cite{H97,Yama}).
This led to a wide application of replica-exchange molecular
dynamics method in the protein folding 
problem \cite{Gar}-\cite{FE03}.

However, REM also has a computational difficulty:
As the number of degrees of freedom of the system increases,
the required number of replicas also greatly increases, whereas 
only a single replica is simulated in MUCA or ST.
This demands a lot of computer power for complex systems.
Our solution to this problem is: Use REM for the weight
factor determinations of MUCA or ST, which is much
simpler than previous iterative methods of weight
determinations, and then perform a long MUCA or ST
production run.
The first example is
the {\it replica-exchange multicanonical algorithm}\index{replica-exchange
multicanonical algorithm} (REMUCA)\index{REMUCA}
\cite{SO3,MSO03,MSO03b}.
In REMUCA,
a short replica-exchange simulation is performed, and the 
multicanonical weight factor is determined by
the multiple-histogram reweighting techniques \cite{FS2,WHAM}.
Another example of such a combination is the
{\it replica-exchange simulated tempering}\index{replica-exchange simulated tempering} 
(REST)\index{REST} \cite{MO4}.
In REST, a short replica-exchange simulation is performed, and
the simulated tempering weight factor is determined by
the multiple-histogram reweighting techniques \cite{FS2,WHAM}.

We have introduced two further extensions of REM,
which we refer to as {\it multicanonical replica-exchange method}\index{multicanonical
replica-exchange method}
(MUCAREM)\index{MUCAREM} \cite{SO3,MSO03,MSO03b} (see also Refs.~\cite{XB00,FYD02}) 
and {\it simulated tempering replica-exchange method}\index{simulated tempering
replica-exchange method}
(STREM)\index{STREM} \cite{STREM} (see also Ref.~\cite{FE03b} for a similar idea).
In MUCAREM, a replica-exchange simulation is
performed
with a small number of replicas each in multicanonical ensemble
of different energy ranges.  In STREM, on the other hand,
a replica-exchange simulation is performed with a small 
number of replicas in ``simulated tempering'' ensemble
of different temperature ranges.

Finally, one is naturally led to a multidimensional 
(or, multivariable) extension of REM, which we refer
to as {\it multidimensional replica-exhcange method}\index{multidimensional replica-exchange
method} (MREM)\index{MREM}
\cite{SKO} (see also Refs.~\cite{Huk2,YP,WBS02,FE03,KH05}).  A special 
realization 
of MREM is {\it replica-exchange umbrella sampling}\index{replica-exchange umbrella sampling} 
(REUS)\index{REUS}
\cite{SKO} and it is particularly useful in free energy calculations
(see also Ref.~\cite{WEK03} for a similar idea).

In this article, we describe the generalized-ensemble algorithms
mentioned above.  Namely, we first review the three familiar methods:
MUCA, ST, and REM.  We then present further extensions of the three methods.

\section{GENERALIZED-ENSEMBLE ALGORITHMS}

\subsection{Multicanonical Algorithm and Simulated Tempering}

Let us consider a system of $N$ atoms of 
mass $m_k$ ($k=1, \cdots, N$)
with their coordinate vectors and
momentum vectors denoted by 
$q \equiv \{{\biq}_1, \cdots, {\biq}_N\}$ and 
$p \equiv \{{\bip}_1, \cdots, {\bip}_N\}$,
respectively.
The Hamiltonian $H(q,p)$ of the system is the sum of the
kinetic energy $K(p)$ and the potential energy $E(q)$:
\begin{equation}
H(q,p) =  K(p) + E(q)~,
\label{eqn1}
\end{equation}
where
\begin{equation}
K(p) =  \sum_{k=1}^N \frac{{\bip_k}^2}{2 m_k}~.
\label{eqn2}
\end{equation}

In the canonical ensemble at temperature $T$ 
each state $x \equiv (q,p)$ with the Hamiltonian $H(q,p)$
is weighted by the Boltzmann factor:
\begin{equation}
W_{\rm B}(x;T) = \exp \left(-\beta H(q,p) \right)~,
\label{eqn3}
\end{equation}
where the inverse temperature $\beta$ is defined by 
$\beta = 1/k_{\rm B} T$ ($k_{\rm B}$ is the Boltzmann constant). 
The average kinetic energy at temperature $T$ is then given by
\begin{equation}
\left< ~K(p)~ \right>_T =  
\left< \sum_{k=1}^N \frac{{\bip_k}^2}{2 m_k} \right>_T 
= \frac{3}{2} N k_{\rm B} T~.
\label{eqn4}
\end{equation}

Because the coordinates $q$ and momenta $p$ are decoupled
in Eq.~(\ref{eqn1}), we can suppress the kinetic energy
part and can write the Boltzmann factor as
\begin{equation}
W_{\rm B}(x;T) = W_{\rm B}(E;T) = \exp (-\beta E)~.
\label{eqn4b}
\end{equation}
The canonical probability distribution of potential energy
$P_{\rm B}(E;T)$ is then given by the product of the
density of states $n(E)$ and the Boltzmann weight factor
$W_{\rm B}(E;T)$:
\begin{equation}
 P_{\rm B}(E;T) \propto n(E) W_{\rm B}(E;T)~.
\label{eqn4c}
\end{equation}
Since $n(E)$ is a rapidly
increasing function and the Boltzmann factor decreases exponentially,
the canonical ensemble yields a bell-shaped distribution
which has a maximum around the average energy at temperature $T$. 
The
conventional MC or MD simulations at constant temperature
are expected to yield $P_{\rm B}(E;T)$.
A MC simulation based on the Metropolis 
algorithm \cite{Metro} is performed with the following 
transition probability from a state $x$
of potential energy $E$ to a state $x^{\prime}$ of 
potential energy $E^{\prime}$:
\begin{equation}
w(x \rightarrow x^{\prime})
= {\rm min} \left(1,\frac{W_{\rm B}(E^{\prime};T)}{W_{\rm B}(E;T)}\right)
= {\rm min} \left(1,\exp \left(-\beta \Delta E \right)\right)~.
\label{eqn4d}
\end{equation}
where
\begin{equation}
\Delta E = E^{\prime} - E~.
\label{eqn4dp}
\end{equation}
A MD simulation, on the other hand, is based on the 
following Newton equations of motion:
\begin{eqnarray}
\dot{\bs{q}_k} &=& \frac{\bs{p}_k}{m_k}~, \\
\dot{\bs{p}_k} &=& 
- \frac{\partial E}{\partial \bs{q}_k} = \bs{f}_k~, 
\label{eqn4e}
\end{eqnarray}
where $\bs{f}_k$ is the force acting on the $k$-th atom
($k = 1, \cdots, N$).
This set of equations actually yield the microcanonical ensemble, and
we have to add a thermostat 
in order to obtain the canonical ensemble at temperature $T$.
Here, we just follow Nos{\'e}'s prescription~\cite{nose84,nosejcp84}, 
and we have
\begin{eqnarray}
\dot{\bs{q}}_k ~&=&~ \frac{\bs{p}_k}{m_k}~, \label{eqn4f1}\\
\dot{\bs{p}}_k ~&=&~ 
- \frac{\partial E}{\partial \bs{q}_k} 
- \frac{\dot{s}}{s}~\bs{p}_k~
= \bs{f}_k - \frac{\dot{s}}{s}~\bs{p}_k~, \label{eqn4f2}\\
\dot{s} ~&=&~ s~\frac{P_s}{Q}~, \label{eqn4f3}\\
\dot{P}_s ~&=&~ 
\sum_{k=1}^{N} \frac{{\bs{p}_k}^{2}}{m_k} - 3N k_{\rm B} T 
~=~3Nk_{\rm B} \left( T(t) - T \right)~, 
\label{eqn4f}
\end{eqnarray}
where $s$ is Nos{\'e}'s scaling parameter, $Q$ is its mass,
$P_s$ is its conjugate momentum, and
the ``instantaneous temperature'' $T(t)$ is defined by
\begin{equation}
T(t) = \frac{1}{3N k_{\rm B}} \sum_{k=1}^{N} \frac{\bs{p}_k(t)^{2}}{m_k} ~.
\label{eqn4g}
\end{equation}
  
However, in practice, it is very difficult to obtain accurate canonical
distributions of complex systems at low temperatures by
conventional MC or MD simulation methods.
This is because simulations at low temperatures tend to get
trapped in one or a few of local-minimum-energy states.

In the multicanonical ensemble \cite{MUCA1,MUCA2}, on the other
hand, each state is weighted by
a non-Boltzmann weight
factor $W_{\rm mu}(E)$ (which we refer to as the {\it multicanonical
weight factor}) so that a uniform potential energy
distribution $P_{\rm mu}(E)$ is obtained:
\begin{equation}
 P_{\rm mu}(E) \propto n(E) W_{\rm mu}(E) \equiv {\rm const}~.
\label{eqn5}
\end{equation}
The flat distribution implies that
a free random walk in the potential energy space is realized
in this ensemble.
This allows the simulation to escape from any local minimum-energy states
and to sample the configurational space much more widely than 
the conventional canonical MC or MD methods.

The definition in Eq.~(\ref{eqn5}) implies that
the multicanonical
weight factor is inversely proportional to the density of
states, and we can write it as follows:
\begin{equation}
W_{\rm mu}(E) \equiv \exp \left[-\beta_0 E_{\rm mu}(E;T_0) \right]
= \frac{1}{n(E)}~,
\label{eqn6}
\end{equation}
where we have chosen an arbitrary reference
temperature, $T_0 = 1/k_{\rm B} \beta_0$, and
the ``{\it multicanonical potential energy}''
is defined by
\begin{equation}
 E_{\rm mu}(E;T_0) \equiv k_{\rm B} T_0 \ln n(E) = T_0 S(E)~.
\label{eqn7}
\end{equation}
Here, $S(E)$ is the entropy in the microcanonical
ensemble.
Since the density of states of the system is usually unknown,
the multicanonical weight factor
has to be determined numerically by iterations of short preliminary
runs \cite{MUCA1,MUCA2}. 

A multicanonical MC simulation
is performed, for instance, with the usual Metropolis criterion \cite{Metro}:
The transition probability of state $x$ with potential energy
$E$ to state $x^{\prime}$ with potential energy $E^{\prime}$ is given by
\begin{equation}
w(x \rightarrow x^{\prime})
= {\rm min} \left(1,\frac{W_{\rm mu}(E^{\prime})}{W_{\rm mu}(E)}\right)
= {\rm min} \left(1,\frac{n(E)}{n(E^{\prime})}\right)
= {\rm min} \left(1,\exp \left( - \beta_0 \Delta E_{\rm mu} \right)\right)~,
\label{eqn8}
\end{equation}
where
\begin{equation}
\Delta E_{\rm mu} = E_{\rm mu}(E^{\prime};T_0) - E_{\rm mu}(E;T_0)~.
\label{eqn9}
\end{equation}
The MD algorithm in the multicanonical ensemble
also naturally follows from Eq.~(\ref{eqn6}), in which the
regular constant temperature MD simulation
(with $T=T_0$) is performed by replacing $E$ by $E_{\rm mu}$
in Eq.~(\ref{eqn4f2})~\cite{HOE96,NNK}:  
\begin{equation}
\dot{\bs{p}}_k ~=~ - \frac{\partial E_{\rm mu}(E;T_0)}{\partial \bs{q}_k}
- \frac{\dot{s}}{s}~\bs{p}_k
~=~ \frac{\partial E_{\rm mu}(E;T_0)}{\partial E}~\bs{f}_k
- \frac{\dot{s}}{s}~\bs{p}_k~.
\label{eqn9a}
\end{equation}
From Eq.~(\ref{eqn7}) this equation can be rewritten as
\begin{equation}
\dot{{\bip}}_k ~=~ \frac{T_0}{T(E)}~{\bif}_k
- \frac{\dot{s}}{s}~\bs{p}_k~.
\label{eqn9b}
\end{equation}
where the following thermodynamic relation gives
the definition of the ``effective temperature''
$T(E)$:
\begin{equation}
\left. \frac{\partial S(E)}{\partial E}\right|_{E=E_a}
~=~ \frac{1}{T(E_a)}~,
\label{eqn9c}
\end{equation}
with
\begin{equation}
E_a ~=~ <E>_{T(E_a)}~.
\label{eqn9d}
\end{equation}

If the exact multicanonical weight factor $W_{\rm mu}(E)$ is known, 
one can calculate the ensemble averages of any physical quantity $A$ 
at any temperature 
$T$ ($= 1/k_{\rm B} \beta$) as follows: 
\begin{equation} 
<A>_T =  \frac{ \displaystyle{ 
\sum_E~ A (E) P_{\rm B}(E;T)} }
{\displaystyle{ \sum_E~ P_{\rm B}(E;T)} } =  \frac{ \displaystyle{ 
\sum_E~ A (E) n(E) \exp(-\beta E) } }
{\displaystyle{ \sum_E~ n(E) \exp(-\beta E) } }~, 
\label{eqn10a}
\end{equation}
where the density of states is given by (see Eq. (\ref{eqn6}))
\begin{equation} 
n(E)= \frac{1}{W_{\rm mu}(E)}~.
\label{eqn10b}
\end{equation}
The summation instead of integration is used
in Eq.~(\ref{eqn10a}), because we often
discretize the potential energy $E$ with step size $\epsilon$
($E = E_i; i=1, 2, \cdots$).
Here, the explicit form of the physical quantity $A$
should be known as a function of potential energy $E$.
For instance, $A(E)=E$ gives the average
potential energy $<E>_T$ as a function of temperature, and 
$A(E) = \beta^2 (E - <E>_T)^2$ gives specific heat.

In general, the multicanonical weight factor $W_{\rm mu}(E)$, or
the density of states $n(E)$, is not 
$a \ priori$ known, and one needs its estimator 
for a numerical simulation. 
This estimator is usually obtained 
from iterations of short trial multicanonical simulations. 
The details of this process
are described, for instance, in Refs.~\cite{MUCA3,OH}.
%
However, the iterative process can be non-trivial and very tedius for
complex systems.

In practice, it is impossible to obtain the ideal
multicanonical weight factor with completely uniform
potential energy distribution.
The question is when to stop the iteration for the
weight factor determination.
Our criterion for a satisfactory weight factor is that 
as long as we do get a random walk in potential energy space,
the probability distribution $P_{\rm mu}(E)$ does not have to
be completely
flat with a tolerance of, say, an order of magnitude deviation.
In such a case, we usually 
perform with this weight 
factor a multicanonical simulation with high statistics
(production run) in order to get even better estimate
of the density of states. 
Let $N_{\rm mu}(E)$ be the histogram
of potential energy distribution 
$P_{\rm mu} (E)$ 
obtained by this production run.
The best estimate of the density of states can then be
given by the single-histogram reweighting
techniques \cite{FS1} as follows (see the proportionality
relation in Eq. (\ref{eqn5})):
\begin{equation} 
n(E)= \displaystyle{\frac{N_{\rm mu}(E)}{W_{\rm mu}(E)}}~.
\label{eqn10c}
\end{equation}
By substituting this quantity into Eq. (\ref{eqn10a}),
one can calculate ensemble averages of physical
quantity $A(E)$ as a function of temperature.
Moreover, ensemble averages of any physical quantity $A$
(including those that cannot be expressed as functions
of potential energy) at any temperature 
$T$ ($= 1/k_{\rm B} \beta$) can now be obtained as long as
one stores the ``trajectory'' of configurations (and $A$)
from the production run.  
Namely, we have
\begin{equation} 
<A>_T  =  \frac{ \displaystyle{ 
\sum_{k=1}^{n_0} A(x(k)) W_{\rm mu}^{-1} (E(x(k)))
\exp \left[-\beta E(x(k)) \right] } }
{\displaystyle{ 
\sum_{k=1}^{n_0} W_{\rm mu}^{-1} (E(x(k))) 
\exp \left[-\beta E(x(k)) \right] } }~, 
\label{eqn10d}
\end{equation}
where $x(k)$ is the configuration at the
$k$-th MC (or MD) step and $n_0$ is the total number 
of configurations stored. 
Note that when $A$ is a function of $E$, Eq.~(\ref{eqn10d}) reduces to
Eq.~(\ref{eqn10a}) where the density of states is given by
Eq.~(\ref{eqn10c}).

Eqs. (\ref{eqn10a}) and (\ref{eqn10d}) or any other equations
which involve summations of
exponential functions often encounter with numerical difficulties
such as overflows.  These can be overcome by using, 
for instance,
the following equation \cite{BergTXT,BergLog}:
For $C=A+B$ (with $A>0$ and $B>0$) we have
\begin{equation}
\begin{array}{rl} 
\ln C &= \ln \left[{\rm max}(A,B) \left(1 +
 \displaystyle{\frac{{\rm min}(A,B)}{{\rm max}(A,B)}} 
 \right) \right]~, \\
 &= {\rm max}(\ln A, \ln B) +
 \ln \left\{1+\exp \left[{\rm min}(\ln A,\ln B) -
 {\rm max}(\ln A,\ln B) \right] \right\}~.
\end{array}
\label{eqn10e}
\end{equation}

We now briefly review the original {\it simulated tempering} (ST)
method \cite{ST1,ST2}.  In this method temperature itself becomes a
dynamical variable, and both the configuration and the temperature are updated
during the simulation with a weight:
\begin{equation}
W_{\rm ST} (E;T) = \exp \left(-\beta E + a(T) \right)~,
\label{Eqn1}
\end{equation}
where the function $a(T)$ is chosen so that the probability distribution
of temperature is flat:
\begin{equation}
P_{\rm ST}(T) = \int dE~ n(E)~ W_{{\rm ST}} (E;T) =
\int dE~ n(E)~ \exp \left(-\beta E + a(T) \right) = {\rm const}~.
\lab{Eqn2}
\end{equation}
Hence, in simulated tempering the {\it temperature} is sampled
uniformly. A free random walk in temperature space
is realized, which in turn
induces a random walk in potential energy space and
allows the simulation to escape from
states of energy local minima.

In the numerical work we discretize the temperature in
$M$ different values, $T_m$ ($m=1, \cdots, M$).  Without loss of
generality we can order the temperature
so that $T_1 < T_2 < \cdots < T_M$.  The lowest temperature
$T_1$ should be sufficiently low so that the simulation can explore the
global-minimum-energy region, and
the highest temperature $T_M$ should be sufficiently high so that
no trapping in an energy-local-minimum state occurs.  The probability
weight factor in Eq.~(\ref{Eqn1}) is now written as
\begin{equation}
W_{\rm ST}(E;T_m) = \exp (-\beta_m E + a_m)~,
\label{Eqn3}
\end{equation}
where $a_m=a(T_m)$ ($m=1, \cdots, M$).
Note that from Eqs.~(\ref{Eqn2}) and (\ref{Eqn3}) we have
\begin{equation}
\exp (-a_m) \propto \int dE~ n(E)~ \exp (- \beta_m E)~.
\label{Eqn4}
\end{equation}
The parameters $a_m$ are therefore ``dimensionless'' Helmholtz free energy
at temperature $T_m$
(i.e., the inverse temperature $\beta_m$ multiplied by
the Helmholtz free energy).
We remark that the density of states $n(E)$ (and hence, the
multicanonical weight factor) and the simulated tempering
weight factor $a_m$ are related by a Laplace transform
\cite{HO96a}.
The knowledge of one implies that of the other, although
in numerical work the inverse Laplace transform of 
Eq. (\ref{Eqn4}) is nontrivial.

Once the parameters $a_m$ are determined and the initial configuration and
the initial temperature $T_m$ are chosen,
a simulated tempering simulation is then realized by alternately
performing the following two steps \cite{ST1,ST2}:
\begin{enumerate}
\item A canonical MC or MD simulation at the fixed temperature $T_m$
(based on Eq. (\ref{eqn4d}) or Eq. (\ref{eqn4e})) is carried out 
for a certain steps.
\item The temperature $T_m$ is updated to the neighboring values
$T_{m \pm 1}$ with the configuration fixed.  The transition probability of
this temperature-updating
process is given by the Metropolis criterion (see Eq.~(\ref{Eqn3})):
\begin{equation}
w(T_m \rightarrow T_{m \pm 1})
= {\rm min}\left(1,\frac{W_{\rm ST}(E;T_{m \pm 1})}{W_{\rm ST}(E;T_m)}\right)
= {\rm min}\left(1,\exp \left( - \Delta \right)\right)~,
\label{Eqn5}
\end{equation}
where
\begin{equation}
\Delta = \left(\beta_{m \pm 1} - \beta_m \right) E
- \left(a_{m \pm 1} - a_m \right)~.
\label{Eqn6}
\end{equation}
\end{enumerate}
Note that in Step 2 we exchange only pairs of 
neighboring temperatures in order to secure sufficiently
large acceptance ratio of temperature updates.

As in multicanonical algorithm, the simulated tempering
parameters $a_m=a(T_m)$ ($m=1, \cdots, M$)
are also determined by iterations of short trial simulations
(see, e.g.,  Refs.~\cite{STrev,IRB1,HO96b} for details).
This process can be non-trivial and very tedius for complex
systems.

After the optimal simulated tempering weight factor is determined,
one performs a long simulated tempering run once.
The canonical expectation value of a physical quantity $A$
at temperature $T_m$ ($m=1, \cdots, M$) can be calculated
by the usual arithmetic mean as follows:
\begin{equation}
<A>_{T_m} = \frac{1}{n_{m}} \sum_{k=1}^{n_{m}}
A\left(x_{m}(k)\right)~,
\label{Eqn7}
\end{equation}
where $x_m(k)$ ($k=1,\cdots,n_m$) are the configurations 
obtained at temperature $T_m$
and $n_{m}$ is the total number of measurements made
at $T=T_m$.
The expectation value at any intermediate temperature
can also be obtained from
Eq.~(\ref{eqn10a}), where the 
density of states is given by
the multiple-histogram reweighting techniques \cite{FS2,WHAM}
as follows.
Let $N_m(E)$ and $n_m$ be respectively
the potential-energy histogram and the total number of
samples obtained at temperature $T_m=1/k_{\rm B} \beta_m$
($m=1, \cdots, M$). 
The best estimate of the density of states is then given by \cite{FS2,WHAM}
\begin{equation}
n(E) = \frac{\dis{\sum_{m=1}^M ~g_m^{-1}~N_m(E)}}
{\dis{\sum_{m=1}^M ~g_m^{-1}~n_m~\exp (f_m-\beta_m E)}}~,
\label{Eqn8a}
\end{equation}
where we have for each $m$ ($=1, \cdots, M$)
\begin{equation}
\exp (-f_m) = \sum_{E} ~n(E)~\exp (-\beta_m E)~.
\label{Eqn8b}
\end{equation}
Here, $g_m = 1 + 2 \tau_m$,
and $\tau_m$ is the integrated
autocorrelation time at temperature $T_m$.
For many systems the quantity $g_m $ can safely 
be set to be a constant in the reweighting 
formulae \cite{WHAM}, and hereafter we set $g_m =1$. 

Note that
Eqs.~(\ref{Eqn8a}) and
(\ref{Eqn8b}) are solved self-consistently
by iteration \cite{FS2,WHAM} to obtain
the density of states $n(E)$ and
the dimensionless Helmholtz free energy $f_m$.
Namely, we
can set all the $f_m$ ($m=1, \cdots, M$) to, e.g., zero initially.
We then use Eq.~(\ref{Eqn8a}) to obtain 
$n(E)$, which is substituted into
Eq.~(\ref{Eqn8b}) to obtain next values of $f_m$, and so on.

Moreover, ensemble averages of any physical quantity $A$
(including those that cannot be expressed as functions
of potential energy) at any temperature 
$T$ ($= 1/k_{\rm B} \beta$) can now be obtained from
the ``trajectory'' of configurations of 
the production 
run.  Namely, we first obtain $f_m$ ($m=1, \cdots, M$) by solving
Eqs. (\ref{Eqn8a}) and (\ref{Eqn8b}) self-consistently, 
and then we have \cite{MSO03}
\begin{equation}
<A>_T  =  \frac{ \displaystyle{ 
\sum_{m=1}^{M} \sum_{k=1}^{n_m} A(x_m(k))  \frac{1}
{\displaystyle{\sum_{\ell=1}^{M} n_{\ell} 
\exp \left[ f_{\ell} - \beta_{\ell} E(x_m(k)) \right]}}
\exp \left[ -\beta E(x_m(k)) \right] }}
{\displaystyle{ 
\sum_{m=1}^{M} \sum_{k=1}^{n_m} \frac{1}
{\displaystyle{\sum_{\ell=1}^{M} n_{\ell} 
\exp \left[ f_{\ell} - \beta_{\ell} E(x_m(k)) \right]}}
\exp \left[ -\beta E(x_m(k)) \right] }}~,
\label{eqn17}
\end{equation}
where $x_m(k)$ ($k=1,\cdots,n_m$) are the configurations 
obtained at temperature $T_m$. \\ 

\subsection{Replica-Exchange Method}

The {\it replica-exchange method} (REM) \cite{RE1}--\cite{RE2}
was developed as an extension of
simulated tempering \cite{RE1} (thus it is also referred to as
{\it parallel tempering} \cite{STrev})
(see, e.g.,  Ref.~\cite{SO} for a detailed
description of the algorithm).
The system for REM consists of 
$M$ {\it non-interacting} copies (or, replicas) 
of the original system in the canonical ensemble
at $M$ different temperatures $T_m$ ($m=1, \cdots, M$).
We arrange the replicas so that there is always
exactly one replica at each temperature.
Then there exists a one-to-one correspondence between replicas
and temperatures; the label $i$ ($i=1, \cdots, M$) for replicas 
is a permutation of 
the label $m$ ($m=1, \cdots, M$) for temperatures,
and vice versa:
\begin{equation}
\left\{
\begin{array}{rl}
i &=~ i(m) ~\equiv~ f(m)~, \cr
m &=~ m(i) ~\equiv~ f^{-1}(i)~,
\end{array}
\right.
\label{eq4b}
\end{equation}
where $f(m)$ is a permutation function of $m$ and
$f^{-1}(i)$ is its inverse.

Let $X = \left\{x_1^{[i(1)]}, \cdots, x_M^{[i(M)]}\right\} 
= \left\{x_{m(1)}^{[1]}, \cdots, x_{m(M)}^{[M]}\right\}$ 
stand for a ``state'' in this generalized ensemble.
Each ``substate'' $x_m^{[i]}$ is specified by the
coordinates $q^{[i]}$ and momenta $p^{[i]}$
of $N$ atoms in replica $i$ at temperature $T_m$:
\begin{equation}
x_m^{[i]} \equiv \left(q^{[i]},p^{[i]}\right)_m~.
\label{eq5}
\end{equation}

Because the replicas are non-interacting, the weight factor for
the state $X$ in
this generalized ensemble is given by
the product of Boltzmann factors for each replica (or at each
temperature):
\begin{equation}
\begin{array}{rl}
W_{\rm REM}(X) 
&= \displaystyle{ \prod_{i=1}^{M}
\exp \left\{- \beta_{m(i)} 
H\left(q^{[i]},p^{[i]}\right) \right\} } 
 = \displaystyle{ \prod_{m=1}^{M}
  \exp \left\{- \beta_m 
H\left(q^{[i(m)]},p^{[i(m)]}\right)
 \right\} }~, \cr
&= \exp \left\{- \dis{\sum_{i=1}^M \beta_{m(i)} 
H\left(q^{[i]},p^{[i]}\right) } \right\}
 = \exp \left\{- \dis{\sum_{m=1}^M \beta_m 
H\left(q^{[i(m)]},p^{[i(m)]}\right) }
 \right\}~,
\end{array}
\label{eq7}
\end{equation}
where $i(m)$ and $m(i)$ are the permutation functions in 
Eq.~(\ref{eq4b}).

We now consider exchanging a pair of replicas in the generalized
ensemble.  Suppose we exchange replicas $i$ and $j$ which are
at temperatures $T_m$ and $T_n$, respectively:  
\begin{equation}
X = \left\{\cdots, x_m^{[i]}, \cdots, x_n^{[j]}, \cdots \right\} 
\longrightarrow \ 
X^{\prime} = \left\{\cdots, x_m^{[j] \prime}, \cdots, x_n^{[i] \prime}, 
\cdots \right\}~. 
\label{eq8}
\end{equation}
Here, $i$, $j$, $m$, and $n$ are related by the permutation
functions in Eq.~(\ref{eq4b}),
and the exchange of replicas introduces a new 
permutation function $f^{\prime}$:
\begin{equation}
\left\{
\begin{array}{rl}
i &= f(m) \longrightarrow j=f^{\prime}(m)~, \cr
j &= f(n) \longrightarrow i=f^{\prime}(n)~. \cr
\end{array}
\right.
\label{eq8c}
\end{equation}

The exchange of replicas can be written in more detail as
\begin{equation}
\left\{
\begin{array}{rl}
x_m^{[i]} \equiv \left(q^{[i]},p^{[i]}\right)_m & \longrightarrow \ 
x_m^{[j] \prime} \equiv \left(q^{[j]},p^{[j] \prime}\right)_m~, \cr
x_n^{[j]} \equiv \left(q^{[j]},p^{[j]}\right)_n & \longrightarrow \ 
x_n^{[i] \prime} \equiv \left(q^{[i]},p^{[i] \prime}\right)_n~,
\end{array}
\right.
\label{eq9}
\end{equation}
where the definitions for $p^{[i] \prime}$ and $p^{[j] \prime}$
will be given below.
We remark that this process is equivalent to exchanging
a pair of temperatures $T_m$ and $T_n$ for the
corresponding replicas $i$ and $j$ as follows:
\begin{equation}
\left\{
\begin{array}{rl}
x_m^{[i]} \equiv \left(q^{[i]},p^{[i]}\right)_m & \longrightarrow \ 
x_n^{[i] \prime} \equiv \left(q^{[i]},p^{[i] \prime}\right)_n~, \cr
x_n^{[j]} \equiv \left(q^{[j]},p^{[j]}\right)_n & \longrightarrow \ 
x_m^{[j] \prime} \equiv \left(q^{[j]},p^{[j] \prime}\right)_m~.
\end{array}
\right.
\label{eq10}
\end{equation}

In the original implementation of the 
{\it replica-exchange method} (REM) \cite{RE1}--\cite{RE2},
Monte Carlo algorithm was used, and only the coordinates $q$
(and the potential energy
function $E(q)$)
had to be taken into account.  
In molecular dynamics algorithm, on the other hand, we also have to
deal with the momenta $p$.
We proposed the following momentum 
assignment in Eq.~(\ref{eq9}) (and in Eq.~(\ref{eq10})) \cite{SO}:
\begin{equation}
\left\{
\begin{array}{rl}
p^{[i] \prime} & \equiv \dis{\sqrt{\frac{T_n}{T_m}}} ~p^{[i]}~, \cr
p^{[j] \prime} & \equiv \dis{\sqrt{\frac{T_m}{T_n}}} ~p^{[j]}~,
\end{array}
\right.
\label{eq11}
\end{equation}
which we believe is the simplest and the most natural.
This assignment means that we just rescale uniformly 
the velocities of all the atoms 
in the replicas by
the square root of the ratio of the two temperatures so that
the temperature condition in Eq.~(\ref{eqn4}) may be satisfied.

In order for this exchange process to converge towards an equilibrium
distribution, it is sufficient to impose the detailed balance
condition on the transition probability $w(X \rightarrow X^{\prime})$:
\begin{equation}
\frac{W_{\rm REM}(X)}{Z} \  w(X \rightarrow X^{\prime})
= \frac{W_{\rm REM}(X^{\prime})}{Z} \  w(X^{\prime} \rightarrow X)~,
\label{eq12}
\end{equation}
where $Z$ is the partition function of the entire system.
From Eqs.~(\ref{eqn1}), (\ref{eqn2}), (\ref{eq7}), (\ref{eq11}), 
and (\ref{eq12}), we have
\begin{equation}
\begin{array}{rl}
\dis{\frac{W_{\rm REM}(X^{\prime})}{W_{\rm REM}(X)}}
&= \exp \left\{ 
- \beta_m \left[K\left(p^{[j] \prime}\right) + E\left(q^{[j]}\right)\right] 
- \beta_n \left[K\left(p^{[i] \prime}\right) + E\left(q^{[i]}\right)\right]
\right. \cr
& \ \ \ \ \ \ \ \ \ \  \left.
+ \beta_m \left[K\left(p^{[i]}\right) + E\left(q^{[i]}\right)\right] 
+ \beta_n \left[K\left(p^{[j]}\right) + E\left(q^{[j]}\right)\right]
\right\}~, \cr
&= \exp \left\{ 
- \beta_m \dis{\frac{T_m}{T_n}} K\left(p^{[j]}\right)
- \beta_n \dis{\frac{T_n}{T_m}} K\left(p^{[i]}\right)
+ \beta_m K\left(p^{[i]}\right)
+ \beta_n K\left(p^{[j]}\right)
\right. \cr
& \ \ \ \ \ \ \ \ \ \  \left.
- \beta_m \left[E\left(q^{[j]}\right)
                - E\left(q^{[i]}\right)\right] 
- \beta_n \left[E\left(q^{[i]}\right)
                - E\left(q^{[j]}\right)\right] 
\right\}~, \cr
&= \exp \left( - \Delta \right)~,
\end{array}
\label{eq13}
\end{equation}
where
\begin{eqnarray}
\Delta &=& \beta_m 
\left(E\left(q^{[j]}\right) - E\left(q^{[i]}\right)\right) 
- \beta_n
\left(E\left(q^{[j]}\right) - E\left(q^{[i]}\right)\right)~,
\label{eqn14a} \\
  &=& \left(\beta_m - \beta_n \right)
\left(E\left(q^{[j]}\right) - E\left(q^{[i]}\right)\right)~, 
\label{eqn14b}
\end{eqnarray}
and $i$, $j$, $m$, and $n$ are related by the permutation
functions in Eq.~(\ref{eq4b}) before the exchange:
\begin{equation}
\left\{
\begin{array}{ll}
i &= f(m)~, \cr
j &= f(n)~.
\end{array}
\right.
\label{eq13b}
\end{equation}
This can be satisfied, for instance, by the usual Metropolis criterion
\cite{Metro} (see also Eqs.~(\ref{eqn4d}), (\ref{eqn8}), and (\ref{Eqn5})):
\begin{equation}
w(X \rightarrow X^{\prime}) \equiv
w\left( x_m^{[i]} ~\left|~ x_n^{[j]} \right. \right) 
= {\rm min}\left(1,\exp \left( - \Delta \right)\right)~,
\label{eq15}
\end{equation}
where in the second expression 
(i.e., $w( x_m^{[i]} | x_n^{[j]} )$) 
we explicitly wrote the
pair of replicas (and temperatures) to be exchanged.
Note that this is exactly the same criterion that was originally
derived for Monte Carlo algorithm \cite{RE1}--\cite{RE2}.

Without loss of generality we can
again assume $T_1 < T_2 < \cdots < T_M$.
A simulation of the 
{\it replica-exchange method} (REM) \cite{RE1}--\cite{RE2}
is then realized by alternately performing the following two
steps:
\begin{enumerate}
\item Each replica in canonical ensemble of the fixed temperature 
is simulated $simultaneously$ and $independently$
for a certain MC or MD steps. 
\item A pair of replicas at neighboring temperatures,
say $x_m^{[i]}$ and $x_{m+1}^{[j]}$, are exchanged
with the probability
$w\left( x_m^{[i]} ~\left|~ x_{m+1}^{[j]} \right. \right)$ 
in Eq.~(\ref{eq15}).
\end{enumerate}
Note that in Step 2 we exchange only pairs of replicas corresponding to
neighboring temperatures, because
the acceptance ratio of the exchange process decreases exponentially
with the difference of the two $\beta$'s (see Eqs.~(\ref{eqn14b})
and (\ref{eq15})).
Note also that whenever a replica exchange is accepted
in Step 2, the permutation functions in Eq.~(\ref{eq4b})
are updated.

The REM simulation is particularly suitable for parallel
computers.  Because one can minimize the amount of information
exchanged among nodes, it is best to assign each replica to
each node (exchanging pairs of temperature values among nodes
is much faster than exchanging coordinates and momenta).
This means that we keep track of the permutation function
$m(i;t)=f^{-1}(i;t)$ in Eq.~(\ref{eq4b}) as a function
of MC or MD step $t$ during the simulation.
After parallel canonical MC or MD simulations for a certain
steps (Step 1), $M/2$ pairs of
replicas corresponding to neighboring temperatures
are simulateneously exchanged (Step 2), and the pairing is alternated 
between the two possible choices, i.e., ($T_1,T_2$), ($T_3,T_4$), $\cdots$
and ($T_2,T_3$), ($T_4,T_5$), $\cdots$.

The major advantage of REM over other generalized-ensemble
methods such as multicanonical algorithm \cite{MUCA1,MUCA2}
and simulated tempering \cite{ST1,ST2}
lies in the fact that the weight factor 
is {\it a priori} known (see Eq.~(\ref{eq7})), while
in the latter algorithms the determination of the
weight factors can be very tedius and time-consuming.
A random walk in ``temperature space'' is
realized for each replica, which in turn induces a random
walk in potential energy space.  This alleviates the problem
of getting trapped in states of energy local minima.
In REM, however, the number of required replicas increases
as the system size $N$ increases (according to $\sqrt N$) \cite{RE1}.
This demands a lot of computer power for complex systems. 

\subsection{Replica-Exchange Multicanonical Algorithm and
Replica-Exchange Simulated Tempering} 

The {\it replica-exchange multicanonical algorithm} (REMUCA) 
\cite{SO3,MSO03,MSO03b} overcomes
both the difficulties of MUCA (the multicanonical weight factor
determination is non-trivial)
and REM (a lot of replicas, or computation time, is required).
In REMUCA we first perform a short REM simulation (with $M$ replicas)
to determine the
multicanonical weight factor and then perform with this weight
factor a regular multicanonical simulation with high statistics.
The first step is accomplished by the multiple-histogram reweighting
techniques \cite{FS2,WHAM}.
Let $N_m(E)$ and $n_m$ be respectively
the potential-energy histogram and the total number of
samples obtained at temperature $T_m$ ($=1/k_{\rm B} \beta_m$) 
of the REM run.
The density of states $n(E)$ is then given by solving 
Eqs.~(\ref{Eqn8a}) and (\ref{Eqn8b}) self-consistently by iteration.

Once the estimate of the density of states is obtained, the
multicanonical weight factor can be directly determined from
Eq.~(\ref{eqn6}) (see also Eq.~(\ref{eqn7})).
Actually, the density of states $n(E)$ and 
the multicanonical potential energy, $E_{\rm mu}(E;T_0)$,
thus determined are only reliable in the following range:
\begin{equation}
E_1 \le E \le E_M~,
\label{eqn29}
\end{equation}
where 
\begin{equation}
\left\{
\begin{array}{rl}
E_1 &=~ <E>_{T_1}~, \\
E_M &=~ <E>_{T_M}~,
\end{array}
\right.
\label{eqn29b}
\end{equation}
and $T_1$ and $T_M$ are respectively the lowest and the highest
temperatures used in the REM run.
Outside this range we extrapolate
the multicanonical potential energy linearly: \cite{SO3}
\begin{equation}
 {\cal E}_{\rm mu}^{\{0\}}(E) \equiv \left\{
   \begin{array}{@{\,}ll}
   \left. \dis{\frac{\partial E_{\rm mu}(E;T_0)}{\partial E}}
        \right|_{E=E_1} (E - E_1)
             + E_{\rm mu}(E_1;T_0)~, &
         \mbox{for $E < E_1$,} \\
         E_{\rm mu}(E;T_0)~, &
         \mbox{for $E_1 \le E \le E_M$,} \\
   \left. \dis{\frac{\partial E_{\rm mu}(E;T_0)}{\partial E}}
        \right|_{E=E_M} (E - E_M)
             + E_{\rm mu}(E_M;T_0)~, &
         \mbox{for $E > E_M$.}
   \end{array}
   \right.
\label{eqn31}
\end{equation}
The multicanonical MC and MD runs  
are then performed respectively with
the Metropolis criterion of Eq.~(\ref{eqn8})
and with the modified Newton equation 
in Eq.~(\ref{eqn9a}), 
in which 
${\cal E}_{\rm mu}^{\{0\}}(E)$ in
Eq.~(\ref{eqn31}) is substituted into $E_{\rm mu}(E;T_0)$.
We expect to obtain a flat potential energy distribution in
the range of Eq.~(\ref{eqn29}).
Finally, the results are analyzed by the single-histogram
reweighting techniques as described in Eq.~(\ref{eqn10c})
(and Eq.~(\ref{eqn10a})).

Some remarks are now in order.
From Eqs.~(\ref{eqn7}), (\ref{eqn9c}), (\ref{eqn9d}),
and (\ref{eqn29b}), Eq.~(\ref{eqn31}) becomes
\begin{equation}
 {\cal E}_{\rm mu}^{\{0\}}(E) = \left\{
   \begin{array}{@{\,}ll}
    \dis{\frac{T_0}{T_1}} (E - E_1) + T_0 S(E_1) = 
    \dis{\frac{T_0}{T_1}} E + {\rm const}~, &
         \mbox{for $E < E_1 \equiv <E>_{T_1}$,} \\
         T_0 S(E)~, &
         \mbox{for $E_1 \le E \le E_M$,} \\
    \dis{\frac{T_0}{T_M}} (E - E_M) + T_0 S(E_M) = 
    \dis{\frac{T_0}{T_M}} E + {\rm const}~, &
         \mbox{for $E > E_M \equiv <E>_{T_M}$.}
   \end{array}
   \right.
\label{eqn31b}
\end{equation}
The Newton equation in Eq.~(\ref{eqn9a}) is then written as
(see Eqs.~(\ref{eqn9b}), (\ref{eqn9c}), and (\ref{eqn9d}))
\begin{equation}
\dot{{\bip}}_k = \left\{
   \begin{array}{@{\,}ll}
   \dis{\frac{T_0}{T_1}}~{\bif}_k
      - \frac{\dot{s}}{s}~\bs{p}_k~, &
         \mbox{for $E < E_1$,} \\
   \dis{\frac{T_0}{T(E)}}~{\bif}_k
      - \frac{\dot{s}}{s}~\bs{p}_k~, &
         \mbox{for $E_1 \le E \le E_M$,} \\
   \dis{\frac{T_0}{T_M}}~{\bif}_k
      - \frac{\dot{s}}{s}~\bs{p}_k~, &
         \mbox{for $E > E_M$.}
   \end{array}
   \right.
\label{eqn31c}
\end{equation}
Because only the product of inverse temperature $\beta$ and
potential energy $E$ enters in the Boltzmann factor
(see Eq.~(\ref{eqn4b})), a rescaling of the potential energy
(or force) by a constant, say $\alpha$, can be considered as
the rescaling of the temperature by $1/\alpha$ \cite{HOE96,Yama}.  
Hence,
our choice of ${\cal E}_{\rm mu}^{\{0\}}(E)$
in Eq.~(\ref{eqn31}) results in a canonical simulation at
$T=T_1$ for $E < E_1$, a multicanonical simulation for
$E_1 \le E \le E_M$, and a canonical simulation at
$T=T_M$ for $E > E_M$.
Note also that the above arguments are independent of
the value of $T_0$, and we
will get the same results, regardless of its value.

For Monte Carlo method, the above statement follows
directly from the following equation.  Namely, our
choice of the multicanonical potential energy in
Eq.~(\ref{eqn31}) gives (by substituting
Eq.~(\ref{eqn31b}) into Eq.~(\ref{eqn6}))
\begin{equation}
W_{\rm mu}(E) = \exp \left[-\beta_0 {\cal E}_{\rm mu}^{\{0\}}(E) \right]
 = \left\{
   \begin{array}{@{\,}ll}
   \dis{\exp \left(-\beta_1 E + {\rm const}\right)}~, &
         \mbox{for $E < E_1$,} \\
   \dis{\frac{1}{n(E)}}~, &
         \mbox{for $E_1 \le E \le E_M$,} \\
   \dis{\exp \left(-\beta_M E + {\rm const}\right)}~, &
         \mbox{for $E > E_M$.}
   \end{array}
   \right.
\label{eqn31d}
\end{equation}

We now present another effective method of the multicanonical
weight factor determination \cite{RevSO}, which is closely related to
REMUCA.
We first perform a short REM simulation as in REMUCA
and calculate $<E>_{T}$ as a function of $T$
by the multiple-histogram
reweighting techniques (see Eqs.~(\ref{Eqn8a}) and (\ref{Eqn8b})).
Let us recall the Newton equation of
Eq.~(\ref{eqn9b}) and the thermodynamic
relation of Eqs.~(\ref{eqn9c}) and (\ref{eqn9d}).
The effective temperature $T(E)$, or the derivative
$\frac{\partial E_{\rm mu}(E;T_0)}{\partial E}$,
can be numerically obtained as the inverse function of
Eq.~(\ref{eqn9d}), 
where the average $<E>_{T(E)}$ has 
been obtained from the results of the REM simulation
by the multiple-histogram reweighting techniques.
Given its derivative, the multicanonical potential
energy can then be obtained by numerical integration 
(see Eqs.~(\ref{eqn7}) and (\ref{eqn9c})): \cite{RevSO}
\begin{equation}
E_{\rm mu}(E;T_0) = 
T_0 \int_{E_1}^{E} \frac{\partial S(E)}{\partial E} dE
= T_0 \int_{E_1}^{E} \frac{dE}{T(E)}~.
\label{EQ10}
\end{equation}
We remark that the same equation was used to obtain
the multicanonical weight factor in Ref.~\cite{H97c},
where $<E>_{T}$ was estimated by simulated
annealing instead of REM.
Essentially the same formulation was also recently
used in Ref.~\cite{TMK03}
to obtain the multicanonical potential energy,
where $<E>_{T}$ was calculated by conventional
canonical simulations.


We finally present the new method which we refer to as the 
{\it replica-exchange simulated tempering} (REST) \cite{MO4}.  
In this method, just as in REMUCA,
we first perform a short REM simulation (with $M$ replicas)
to determine the simulated tempering
weight factor and then perform with this weight
factor a regular ST simulation with high statistics.
The first step is accomplished by 
the multiple-histogram reweighting
techniques \cite{FS2,WHAM}, which give
the dimensionless Helmholtz free energy $f_m$ (see Eqs.~(\ref{Eqn8a})
and (\ref{Eqn8b})).

Once the estimate of the dimensionless Helmholtz free energy $f_m$ are
obtained, the simulated tempering 
weight factor can be directly determined by using
Eq.~(\ref{Eqn3}) where we set $a_m = f_m$ (compare Eq.~(\ref{Eqn4})
with Eq.~(\ref{Eqn8b})).
A long simulated tempering run is then performed with this
weight factor.  
Let $N_m(E)$ and $n_m$ be respectively
the potential-energy histogram and the total number of
samples obtained at temperature $T_m$ ($=1/k_{\rm B} \beta_m$) from this
simulated tempering run.  The multiple-histogram
reweighting techniques of Eqs.~(\ref{Eqn8a}) and (\ref{Eqn8b}) can be used
again to obtain the best estimate of the density of states
$n(E)$.
The expectation value of a physical quantity $A$
at any temperature $T~(= 1/k_{\rm B} \beta)$ is then calculated from
Eq.~(\ref{eqn10a}).

The formulations of REMUCA and REST are simple and straightforward, but
the numerical improvement is great, because the weight factor
determination for MUCA and ST becomes very difficult
by the usual iterative processes for complex systems.

\subsection{Multicanonical Replica-Exchange Method and
Simulated Tempering Replica-Exchange Method}

In the previous subsection we presented REMUCA, 
which uses a short REM run for the determination 
of the multicanonical weight factor. 
Here, we present two modifications of REM and refer the new methods as 
multicanonical replica-exchange method (MUCAREM) \cite{SO3,MSO03,MSO03b}
and simulated tempering replica-exchange method (STREM)
\cite{STREM}. 
In MUCAREM  the production run is 
a REM simulation with a few replicas
not in the canonical ensemble but
in the multicanonical ensemble, i.e.,
different replicas perform MUCA simulations with
different energy ranges.  Likewise in STREM
the production run is a REM simulation with a few replicas
that performs ST simulations with different temperature
ranges.
While MUCA and ST simulations are usually based on local
updates, a replica-exchange process can be considered to be
a global update, and global updates enhance the sampling
further.

We first describe MUCAREM.
Let ${\cal M}$ be the number of replicas.  Here, each replica
is in one-to-one correspondence not with temperature but with
multicanonical weight factors of different energy range.
Note that because multicanonical simulations cover much wider energy
ranges than regular canonical simulations, the number of
required replicas for the production run of MUCAREM is
much less than that for the regular REM (${\cal M} \ll M$).
The weight factor for this generalized ensemble
is now given by (see Eq. (\ref{eq7}))
\begin{equation}
W_{\rm MUCAREM} (X) = \displaystyle{ \prod_{i=1}^{{\cal M}} 
W_{\rm mu}^{\{m(i)\}}} \left( E \left(x_{m(i)}^{[i]} \right) \right) 
= \displaystyle{ \prod_{m=1}^{{\cal M}} 
W_{\rm mu}^{\{m\}}} \left( E \left( x_m^{[i(m)]} \right) \right)~,
\end{equation}
where we prepare the multicanonical weight factor (and the
density of states) separately for $m$ regions (see Eq.~(\ref{eqn6})):
\begin{equation}
W_{\rm mu}^{\{m\}} \left( E \left(x_m^{[i]} \right) \right) = 
\exp \left[- \beta_m {\cal E}_{\rm mu}^{\{m\}}
\left( E \left(x_m^{[i]} \right) \right) \right]
\equiv \frac{1}{n^{\{m\}}\left(E\left(x_m^{[i]}\right)\right)}~.
\label{eqn23}
\end{equation}
Here, we have introduced ${\cal M}$ arbitrary 
reference temperatures
$T_m = 1/k_{\rm B} \beta_m$ ($m = 1, \cdots, {\cal M}$), but
the final results will be independent of the values of $T_m$, 
as one can see from the second equality in Eq.~(\ref{eqn23})
(these arbitrary 
temperatures are necessary only for MD simulations).
   
Each multicanonical weight factor 
$W_{\rm mu}^{\{m\}}(E)$, or the density of states
$n^{\{m\}}(E)$, 
is defined as follows. 
For each $m$ ($m=1,\cdots, {\cal M}$), 
we assign a pair of temperatures ($T_{\rm L}^{\{m\}},T_{\rm H}^{\{m\}}$). 
Here, we assume that $T_{\rm L}^{\{m\}} < T_{\rm H}^{\{m\}}$ and arrange the 
temperatures so that 
the neighboring regions covered by the pairs have sufficient overlaps. 
Without loss of generality we can assume 
$T_{\rm L}^{\{1\}} < \cdots < T_{\rm L}^{\{{\cal M}\}}$ and
$T_{\rm H}^{\{1\}} < \cdots < T_{\rm H}^{\{{\cal M}\}}$. 
We define the following quantities:
\begin{equation}
\left\{
\begin{array}{rl}
E_{\rm L}^{\{m\}} &=~ <E>_{{T_{\rm L}}^{\{m\}}}~, \\
E_{\rm H}^{\{m\}} &=~ <E>_{{T_{\rm H}}^{\{m\}}}~,~~ \mbox{($m=1,\cdots,{\cal 
M}$)}~.
\end{array}
\right.
\end{equation}

Suppose that the multicanonical weight factor $W_{\rm mu}(E)$
(or equivalently, the multicanonical potential energy
$E_{\rm mu}(E;T_0)$ in Eq. (\ref{eqn7}))
has been obtained as in REMUCA or by any other methods
in the entire energy range of interest ($E_{\rm L}^{\{1\}} < E < 
E_{\rm H}^{\{{\cal M}\}}$). 
We then have for each $m$ ($m=1,\cdots, {\cal M}$)
the following multicanonical potential energies
(see Eq. (\ref{eqn31})): \cite{SO3}
\begin{equation}
 {\cal E}_{\rm mu}^{\{m\}}(E) = \left\{
   \begin{array}{@{\,}ll}
   \dis{\frac{\partial E_{\rm mu}
   (E_{\rm L}^{\{m\}};T_m)}{\partial E}}
        \left( E - E_{\rm L}^{\{m\}} \right)
        + E_{\rm mu}&\left(E_{\rm L}^{\{m\}};T_m \right)~, 
         \mbox{for $E < E_{\rm L}^{\{m\}}$,} \\
         E_{\rm mu}(E;T_m)~, & ~~~~~~~~
         \mbox{for $E_{\rm L}^{\{m\}} \le E \le E_{\rm H}^{\{m\}}$,} \\
   \dis{\frac{\partial E_{\rm mu}
   (E_{\rm H}^{\{m\}};T_m)}{\partial E}}
       \left( E - E_{\rm H}^{\{m\}} \right)
        + E_{\rm mu}&\left(E_{\rm H}^{\{m\}};T_m \right)~, 
         \mbox{for $E > E_{\rm H}^{\{m\}}$.}
   \end{array}
   \right. 
\label{eqn33}
\end{equation}
 
Finally, a MUCAREM simulation 
is realized by alternately performing the following two steps.
\begin{enumerate}
\item Each replica of the fixed multicanonical ensemble is simulated 
$simultaneously$ and $independently$ for a certain MC or MD steps.
\item A pair of replicas, say $i$ and $j$, which are in neighboring 
multicanonical ensembles, say $m$-th and $(m+1)$-th, 
respectively, are exchanged:
$X = \left\{\cdots, x_m^{[i]}, \cdots, x_{m+1}^{[j]}, \cdots \right\}
\longrightarrow \
X^{\prime} = \left\{\cdots, x_m^{[j]}, \cdots, x_{m+1}^{[i]},
\cdots \right\}$.
The transition probability of this replica exchange is 
given by the Metropolis criterion:
\begin{equation}
w(X \rightarrow X^{\prime}) 
= {\rm min}\left(1,\exp \left( - \Delta \right)\right)~,
\label{eqn28}
\end{equation}
where we now have (see Eq.~(\ref{eqn14a})) \cite{SO3}
\begin{equation}
\begin{array}{rl} 
\Delta &= \beta_{m}
\left\{{\cal E}_{\rm mu}^{\{m\}}\left(E\left(q^{[j]}\right)\right) -
{\cal E}_{\rm mu}^{\{m\}}\left(E\left(q^{[i]}\right)\right)\right\} \\
&-~ \beta_{m+1}
\left\{{\cal E}_{\rm mu}^{\{m+1\}}\left(E\left(q^{[j]}\right)\right) -
{\cal E}_{\rm mu}^{\{m+1\}}\left(E\left(q^{[i]}\right)\right)\right\}~.
\end{array}
\label{Eqn21}
\end{equation}
Here, $E\left(q^{[i]}\right)$ and $E\left(q^{[j]}\right)$ are the potential energy 
of the $i$-th replica and the $j$-th replica, respectively.
\end{enumerate}

Note that in Eq. (\ref{Eqn21}) we need to newly evaluate the 
multicanonical
potential energy, ${\cal E}_{\rm mu}^{\{m\}}(E(q^{[j]}))$ and
${\cal E}_{\rm mu}^{\{m+1\}}(E(q^{[i]}))$, because 
${\cal E}_{\rm mu}^{\{m\}}(E)$ and
${\cal E}_{\rm mu}^{\{n\}}(E)$ are, in
general, different functions for $m \ne n$. 

In this algorithm, the $m$-th multicanonical ensemble actually 
results in a 
canonical simulation at  
$T= T_{\rm L}^{\{m\}}$  
for $E < E_{\rm L}^{\{m\}}$, a multicanonical simulation for 
$E_{\rm L}^{\{m\}} \le E \le E_{\rm H}^{\{m\}}$, and 
a canonical simulation at 
$T=T_{\rm H}^{\{m\}}$ 
for $E > E_{\rm H}^{\{m\}}$, 
while the replica-exchange process samples states of 
the whole energy range 
($E_{\rm L}^{\{1\}} \le E \le E_{\rm H}^{\{{\cal M}\}}$).

For obtaining the canonical distributions at any
intermediate temperature $T$,
the multiple-histogram reweighting techniques \cite{FS2,WHAM}
are again used.
Let $N_m(E)$ and $n_m$ be respectively
the potential-energy histogram and the total number of
samples obtained 
with the multicanonical weight factor
$W_{\rm mu}^{\{m\}}(E)$
($m = 1, \cdots, {\cal M}$).
The expectation value
of a physical quantity $A$ 
at any temperature $T$ ($=1/k_{\rm B} \beta$)
is then obtained from Eq.~(\ref{eqn10a}),
where the best estimate of the density of states is obtained by
solving the WHAM equations,
which now read \cite{SO3}
\begin{equation}
n(E) = \frac{  \displaystyle{ \sum_{m=1}^{\cal M} N_m (E)} }
{ \displaystyle{ \sum_{m=1}^{\cal M} n_m \exp(f_m) W_{\rm 
mu}^{\{m\}}(E) } }
= \frac{\dis{\sum_{m=1}^{\cal M} ~N_m(E)}}
{\dis{\sum_{m=1}^{\cal M} 
~n_m~\exp \left(f_m-\beta_m {\cal E}_{\rm mu}^{\{m\}}(E)\right)}}~,
\label{eqn30}
\end{equation}
and for each $m$ ($=1, \cdots, {\cal M}$)
\begin{equation}
\exp(-f_m) = \sum_E n(E) W_{\rm mu}^{\{m\}}(E)
= \sum_{E} ~n(E)~\exp \left(-\beta_m {\cal E}_{\rm mu}^{\{m\}}(E)\right)~.
\label{Eqn31}
\end{equation}
Note that $W_{\rm mu}^{\{m\}}(E)$ is used 
instead of the Boltzmann factor $\exp (- \beta_m E)$ in
Eqs. (\ref{Eqn8a}) and (\ref{Eqn8b}). 

Moreover, ensemble averages of any physical quantity $A$
(including those that cannot be expressed as functions
of potential energy) at any temperature 
$T$ ($= 1/k_{\rm B} \beta$) can now be obtained from
the ``trajectory'' of configurations of 
the production 
run.  Namely, we first obtain $f_m$ ($m=1, \cdots, {\cal M}$) by solving
Eqs. (\ref{eqn30}) and (\ref{Eqn31}) self-consistently, 
and then we have \cite{MSO03}
\begin{equation} 
<A>_T  =  \frac{ \displaystyle{ 
\sum_{m=1}^{\cal M} \sum_{k=1}^{n_m} A(x_m(k))  \frac{1}
{\displaystyle{\sum_{\ell=1}^{\cal M} n_{\ell} 
\exp(f_{\ell}) W_{\rm mu}^{\{\ell \}}(E(x_m(k)))  }}
\exp \left[-\beta E(x_m(k)) \right] }}
{\displaystyle{ 
\sum_{m=1}^{\cal M} \sum_{k=1}^{n_m} \frac{1}
{\displaystyle{\sum_{\ell=1}^{\cal M} n_{\ell} 
\exp(f_{\ell}) W_{\rm mu}^{\{\ell \}}(E(x_m(k)))  }}
\exp \left[-\beta E(x_m(k)) \right] }}~,
\label{eqn32}
\end{equation}
where the trajectories $x_m(k)$ ($k=1,\cdots,n_m$) are 
taken from each multicanonical simulation 
with the multicanonical weight factor
$W_{\rm mu}^{\{m\}}(E)$ ($m=1, \cdots ,{\cal M}$) separately. \\
are 

As seen above, both REMUCA and MUCAREM can be used to obtain the
multicanonical weight factor, or the density of states,
for the entire potential energy range of interest.
For complex systems, however, a single REMUCA or MUCAREM
simulation is often insufficient.
In such cases we can iterate MUCA (in REMUCA) and/or 
MUCAREM simulations
in which the estimate of the multicanonical weight factor
is updated
by the single- and/or multiple-histogram reweighting techniques,
respectively.


To be more specific, this iterative process can be 
summarized as follows.
The REMUCA production run corresponds to 
a MUCA simulation with the weight factor $W_{\rm mu}(E)$. 
The new estimate of the density of states can be obtained
by the single-histogram reweighting techniques 
of Eq. (\ref{eqn10c}). 
On the other hand, from the MUCAREM production run, 
the improved density of states can be obtained
by the multiple-histogram reweighting
techniques of Eqs. (\ref{eqn30}) and (\ref{Eqn31}).

The improved density of states thus obtained 
leads to a new multicanonical
weight factor (see Eq. (\ref{eqn6})).
The next iteration can be either a MUCA production run (as in REMUCA) or
MUCAREM production run.  The results of this production run may yield
an optimal multicanonical weight factor that yields
a sufficiently flat energy distribution for the entire
energy range of interest.  If not, we can repeat the 
above process by obtaining the third estimate of the multicanonical
weight factor either by a MUCA production run (as in REMUCA) or
by a MUCAREM production run, and so on.

We remark that as the estimate of the multicanonical weight factor becomes
more accurate, one is required to have a less number of
replicas for a successful MUCAREM simulation, because each replica
will have a flat energy distribution for a wider energy range.  Hence, 
for a large, complex system, it is often more efficient to
first try MUCAREM and iteratively reduce the number of replicas so
that eventually one needs only one or a few replicas (instead of trying
REMUCA directly from the beginning and iterating MUCA simulations). \\
   

We now describe the simulated tempering replica-exchange method
(STREM) \cite{STREM}.
Suppose that the simulated tempering weight factor $W_{ST}(E;T_n)$
(or equivalently, the dimensionless Helmholtz free energy
$a_{n}$ in Eq. (\ref{Eqn3}))
has been obtained as in REST or by any other methods
in the entire temperature range of interest ($T_1 \le T_n \le 
T_M$). 
We divide the overlapping temperature ranges into
${\cal M}$ regions (${\cal M} \ll M$).
Suppose each temperature range $m$ has ${\cal N}_m$ temperatures:
$T_k^{\{m\}}$ ($k=1,\cdots,{\cal N}_m$) for
$m=1,\cdots,{\cal M}$.
We assign each temperature range to a replica;
each replica $i$
is in one-to-one correspondence with a
different temperature range $m$ of ST run, where
$T_1^{\{m\}} \le T_k^{\{m\}} \le T_{{\cal N}_m}^{\{m\}}$
($k=1,\cdots,{\cal N}_m$).
We then introduce the replica-exchange process
between neighboring temperature ranges.
This works when we allow sufficient overlaps between
the temperature regions.

A STREM simulation is then realized by alternately 
performing the following two steps. \cite{STREM}
\begin{enumerate}
\item Each replica 
performs a ST simulation within the fixed temperature range
$simultaneously$ and $independently$ for a certain MC or MD steps.
\item A pair of replicas, say $i$ and $j$, which are at, say
$T=T_k^{\{m\}}$ and $T=T_{\ell}^{\{m+1\}}$, in neighboring 
temperature ranges, say $m$-th and $(m+1)$-th, 
respectively, are exchanged:
$X = \left\{\cdots, x_k^{[i]}, \cdots, x_{\ell}^{[j]}, \cdots \right\}
\longrightarrow \
X^{\prime} = \left\{\cdots, x_k^{[j]}, \cdots, x_{\ell}^{[i]},
\cdots \right\}$.
The transition probability of this replica exchange is 
given by the Metropolis criterion:
\begin{equation}
w(X \rightarrow X^{\prime})
= {\rm min}\left(1,\exp \left( - \Delta \right)\right)~,
\label{Eqn22}
\end{equation}
where
\begin{equation}
\Delta \equiv \left(\beta_k^{\{m\}} - \beta_{\ell}^{\{m+1\}} \right)
              \left(E\left(q^{[j]}\right)
                  - E\left(q^{[i]}\right)\right)~. 
\label{Eqn23}
\end{equation}

\end{enumerate}

While in MUCAREM each replica performs a random walk
in multicanonical ensemble of finite energy range,
in STREM each replica performs a random walk by
simulated tempering of finite temperature range.
These ``local'' random walks are made ``global'' to cover the
entire energy range of interest by the replica-exchange
process.

\subsection{Multidimensional Replica-Exchange Method}

We now present our multidimensional extension of REM, which
we refer to as {\it multidimensional replica-exchange method}
(MREM) \cite{SKO}.
The crucial observation that led to the new algorithm is:  
As long as we have $M$ {\it non-interacting}
replicas of the original system, the Hamiltonian 
$H(q,p)$ of the system does not have to be identical
among the replicas and it can depend on a parameter
with different parameter values for different replicas.
Namely, we can write the Hamiltonian for the $i$-th
replica at temperature $T_m$ as
\begin{equation}
H_m (q^{[i]},p^{[i]}) =  K(p^{[i]}) + E_{\lambda_m} (q^{[i]})~,
\label{Eqn16}
\end{equation}
where the potential energy $E_{\lambda_m}$ depends on a
parameter $\lambda_m$ and can be written as
\begin{equation}
E_{\lambda_m} (q^{[i]}) = E_0 (q^{[i]}) + \lambda_m V(q^{[i]})~.
\label{Eqn16p}
\end{equation}
This expression for the potential energy is often used in
simulations.
For instance, in umbrella sampling \cite{US}, $E_0(q)$ and
$V(q)$ can be respectively taken as the original potential
energy and the ``biasing'' potential energy with the
coupling parameter $\lambda_m$.  In simulations of spin 
systems, on the other hand, 
$E_0(q)$ and $V(q)$ (here, $q$ stands for spins)
can be respectively considered as the
zero-field term and the magnetization term coupled with
the external field $\lambda_m$. 

While replica $i$ and temperature
$T_m$ are in one-to-one correspondence
in the original REM,
replica $i$ and ``parameter set''
$\Lambda_m \equiv (T_m,\lambda_m)$ are in one-to-one
correspondence in the new algorithm.
Hence, the present algorithm can be considered as a
multidimensional extension of the original replica-exchange
method where the ``parameter space'' is one-dimensional 
(i.e., $\Lambda_m = T_m$).
Because the replicas are non-interacting, the weight factor 
for the state $X$ in this new
generalized ensemble is again given by the product
of Boltzmann factors for each replica (see Eq.~(\ref{eq7})):
\begin{equation}
\begin{array}{rl}
W_{\rm MREM}(X) &= \exp \left\{- \dis{\sum_{i=1}^M \beta_{m(i)} 
H_{m(i)}\left(q^{[i]},p^{[i]}\right) } \right\}~, \cr
 &= \exp \left\{- \dis{\sum_{m=1}^M \beta_m 
H_m\left(q^{[i(m)]},p^{[i(m)]}\right) }
 \right\}~,
\end{array}
\label{Eqn19}
\end{equation}
where $i(m)$ and $m(i)$ are the permutation functions in 
Eq.~(\ref{eq4b}).
Then the same derivation
that led to the original replica-exchange
criterion follows, and the
transition probability of replica exchange is
given by Eq.~(\ref{eq15}), where
we now have (see Eq.~(\ref{eqn14a})) \cite{SKO}
\begin{equation}
\Delta = \beta_m 
\left(E_{\lambda_m}\left(q^{[j]}\right) - 
E_{\lambda_m}\left(q^{[i]}\right)\right) 
- \beta_n
\left(E_{\lambda_n}\left(q^{[j]}\right) - 
E_{\lambda_n}\left(q^{[i]}\right)\right)~.
\label{eqn21}
\end{equation}
Here, $E_{\lambda_m}$ and $E_{\lambda_n}$ are the
total potential energies (see Eq.~(\ref{Eqn16p})).
Note that we need to newly evaluate the potential
energy for exchanged coordinates,
$E_{\lambda_m} (q^{[j]})$ and $E_{\lambda_n} (q^{[i]})$,
because $E_{\lambda_m}$ and $E_{\lambda_n}$ are in general
different functions.

For obtaining the canonical distributions,
the multiple-histogram reweighting techniques \cite{FS2,WHAM}
are particularly suitable.
Suppose we have made a single run of the present
replica-exchange simulation with $M$ replicas that correspond
to $M$ different parameter sets
$\Lambda_m \equiv (T_m,\lambda_m)$ ($m=1, \cdots, M$).
Let $N_m(E_0,V)$ and $n_m$
be respectively 
the potential-energy histogram and the total number of
samples obtained for the $m$-th parameter set
$\Lambda_m$.
The WHAM equations that yield the canonical
probability distribution 
$P_{T,\lambda} (E_0,V)=n(E_0,V)\exp(-\beta E_\lambda)$
with any
potential-energy parameter value $\lambda$ at
any temperature $T=1/k_{\rm B} \beta$
are then given by \cite{SKO}
\begin{equation}
n(E_0,V)
= \frac{\dis{\sum_{m=1}^M N_m(E_0,V)}} 
{\dis{\sum_{m=1}^M n_{m}~\exp \left(f_m-\beta_m 
E_{\lambda_m}\right)}}~,
\label{eqn19}
\end{equation}
and for each $m$ ($=1, \cdots, M$)
\begin{equation}
\exp (-f_m) = \sum_{E_0,V} n(E_0,V) \exp \left(-\beta_m E_{\lambda_m}\right)~.
\label{eqn20}
\end{equation}
Here, $n(E_0,V)$ is the generalized density of states.
Note that $n(E_0,V)$ is independent of the parameter sets
$\Lambda_m \equiv (T_m,\lambda_m)$ ($m=1, \cdots, M$).
The density of states
$n(E_0,V)$ and the ``dimensionless''
Helmholtz free energy $f_m$ in Eqs.~(\ref{eqn19}) and
(\ref{eqn20}) are solved self-consistently
by iteration.

We can use MREM for
free energy calculations.  We first describe the free-energy
perturbation case.  The potential energy is given by
\begin{equation}
E_{\lambda} (q) = E_I (q) + \lambda \left(E_F (q) - E_I (q)\right)~,
\label{eqn22}
\end{equation}
where $E_I$ and $E_F$ are the potential energy for
a ``wild-type'' molecule and a ``mutated''
molecule, respectively.  Note that this equation has the same
form as Eq.~(\ref{Eqn16p}).

Our replica-exchange simulation is performed for $M$ replicas
with $M$ different values of the parameters
$\Lambda_m = (T_m,\lambda_m)$.
Since $E_{\lambda = 0} (q) = E_I (q)$ and
$E_{\lambda = 1} (q) = E_F (q)$, we should choose enough
$\lambda_m$ values distributed in the range between 0 and 1
so that we may have sufficient acceptance of replica exchange.
From the simulation, $M$ histograms $N_m (E_I,E_F-E_I)$, or
equivalently $N_m(E_I,E_F)$, are obtained.  The Helmholtz
free energy difference of ``mutation'' at 
temperature $T$ $(=1/k_{\rm B}\beta)$,
$\Delta F \equiv F_{\lambda = 1} - F_{\lambda = 0}$, can then
be calculated from 
\begin{equation}
\exp (-\beta \Delta F) = \frac{Z_{T,\lambda=1}} 
{Z_{T,\lambda=0}} =  
\frac{\dis{\sum_{E_I,E_F}
P_{T,\lambda=1} (E_I,E_F)}}
{\dis{\sum_{E_I,E_F}
P_{T,\lambda=0} (E_I,E_F)}} ~,
\label{Eqn24}
\end{equation}
where $P_{T,\lambda} (E_I,E_F) = n(E_I,E_F) \exp \left(-\beta E_{\lambda}\right)$ 
are obtained from the WHAM
equations of Eqs.~(\ref{eqn19}) and (\ref{eqn20}).

We now describe another free energy calculations based on
MREM applied to umbrella sampling \cite{US},
which we refer to as 
{\it replica-exchange umbrella sampling} (REUS).
The potential energy is a generalization of Eq.~(\ref{Eqn16p})
and is given by
\begin{equation}
E_{\bil} (q) = E_0 (q) + \sum_{\ell = 1}^L
\lambda^{(\ell)} V_{\ell} (q)~,
\label{Eqn25}
\end{equation}
where $E_0(q)$ is the original unbiased potential, 
$V_{\ell}(q)$ ($\ell =1, \cdots, L$) are the 
biasing (umbrella) potentials, and $\lambda^{(\ell)}$ are the
corresponding coupling constants
($\bil = (\lambda^{(1)}, \cdots, \lambda^{(L)})$).
Introducing a ``reaction coordinate'' $\xi$,
the umbrella potentials are usually written as harmonic
restraints:
\begin{equation}
V_{\ell} (q) = k_{\ell} \left( \xi (q) - d_{\ell} \right)^2~,
~(\ell =1, \cdots, L)~,
\label{Eqn26}
\end{equation}
where $d_{\ell}$ are the midpoints and $k_{\ell}$ are the
strengths of the restraining potentials.
We prepare $M$ replicas with $M$
different values of the parameters
$\biL_m = (T_m,\bil_m)$, and the replica-exchange
simulation is performed.  Since the umbrella potentials
$V_{\ell} (q)$ in Eq.~(\ref{Eqn26})
are all functions of the reaction coordinate
$\xi$ only, we can take the histogram
$N_m (E_0,\xi)$ instead of
$N_m (E_0,V_1, \cdots, V_L)$.
The WHAM equations of
Eqs.~(\ref{eqn19}) and (\ref{eqn20}) can then be written as \cite{SKO}
\begin{equation}
n(E_0,\xi)
= \frac{\dis{\sum_{m=1}^M ~N_m(E_0,\xi)}} 
{\dis{\sum_{m=1}^M n_{m}~\exp \left(f_m-\beta_m E_{\bil_m}\right)}}~
\label{Eqn27}
\end{equation}
and for each $m$ ($=1, \cdots, M$)
\begin{equation}
\exp (-f_m) = \sum_{E_0,\xi} n(E_0,\xi) \exp \left(-\beta_m E_{\bil_m}\right)~.
\label{Eqn28}
\end{equation}
The expectation value of a physical quantity $A$ 
with any
potential-energy parameter value $\bil$ at
any temperature $T$ ($=1/k_{\rm B} \beta$) is now
given by
\begin{equation}
<A>_{T,\bil} \ = \frac{\dis{\sum_{E_0,\xi}
A(E_0,\xi) P_{T,\bil} (E_0,\xi)}}
{\dis{\sum_{E_0,\xi} P_{T,\bil} (E_0,\xi)}}~,
\label{Eqn29}
\end{equation}
where $P_{T,\bil} (E_0,\xi) = n(E_0,\xi) \exp \left(-\beta E_{\bil}\right)$
is obtained from the WHAM
equations of Eqs.~(\ref{Eqn27}) and (\ref{Eqn28}).

The potential of mean force (PMF), or free energy as a function of
the reaction coordinate, of the original, unbiased system 
at temperature $T$ is given by
\begin{equation}
{\cal W}_{T,\bil = \{0\}} (\xi) = - k_{\rm B} T \ln
\left[ \sum_{E_0} P_{T,\bil = \{0\}} (E_0,\xi) \right]~,
\label{Eqn30}
\end{equation}
where $\{0\} = (0, \cdots, 0)$.

We now present two examples of realization of REUS.
In the first example, we use only one temperature, $T$, and 
$L$ umbrella potentials.
We prepare replicas so that the potential energy for each
replica includes exactly one umbrella potential
(here, we have $M = L$).
Namely, in Eq.~(\ref{Eqn25}) for $\bil = \bil_m$ we set
\begin{equation}
\lambda^{(\ell)}_m = \delta_{\ell,m}~,
\label{Eqnn31}
\end{equation}
where $\delta_{k,l}$ is Kronecker's delta function, and
we have
\begin{equation}
E_{\bil_m} (q^{[i]}) = E_0 (q^{[i]}) + V_m (q^{[i]})~.
\label{Eqn32}
\end{equation}
We exchange 
replicas corresponding to ``neighboring'' umbrella potentials,
$V_{m}$ and $V_{m+1}$.
The acceptance criterion for replica exchange is given
by Eq.~(\ref{eq15}), where Eq.~(\ref{eqn21}) now reads
(with the fixed inverse temperature $\beta = 1/k_{\rm B} T$) \cite{SKO}
\begin{equation}
\Delta = \beta 
\left(V_m\left(q^{[j]}\right) - 
      V_m\left(q^{[i]}\right) -
      V_{m+1}\left(q^{[j]}\right) + 
      V_{m+1}\left(q^{[i]}\right)\right)~,
\label{Eqn33}
\end{equation}
where replicas $i$ and $j$ respectively have umbrella potentials
$V_m$ and $V_{m+1}$ before the exchange.

In the second example, we prepare $N_T$ temperatures
and $L$ umbrella potentials,
which makes the total
number of replicas $M=N_T \times L$.
We can introduce the following re-labeling for the parameters
that characterize the replicas:
\begin{equation}
\begin{array}{rl}
\biL_m = (T_m,\bil_m) & \longrightarrow
\ \biL_{I,J} = (T_I,\bil_J)~. \cr
(m=1, \cdots, M) & \ \ \ \ \ \ \ \ \ \ (I=1, \cdots, N_T,~J=1, \cdots, L)
\end{array}
\label{Eqn34}
\end{equation}
The potential energy is given by Eq.~(\ref{Eqn32})
with the replacement: $m \rightarrow J$.
We perform the following replica-exchange processes alternately:
\begin{enumerate}
\item Exchange pairs of replicas corresponding to neighboring temperatures,
$T_I$ and $T_{I+1}$ 
(i.e., exchange replicas $i$ and $j$ that
respectively correspond to parameters
$\biL_{I,J}$ and $\biL_{I+1,J}$).
(We refer to this process as $T$-exchange.)
\item Exchange pairs of replicas corresponding to 
``neighboring'' umbrella potentials,
$V_J$ and $V_{J+1}$ 
(i.e., exchange replicas $i$ and $j$ that
respectively correspond to parameters
$\biL_{I,J}$ and $\biL_{I,J+1}$).
(We refer to this process as $\lambda$-exchange.)
\end{enumerate}
The acceptance criterion for these 
replica exchanges is given by 
Eq.~(\ref{eq15}), where Eq.~(\ref{eqn21}) now reads \cite{SKO}
\begin{equation}
\Delta = \left(\beta_{I} - \beta_{I+1} \right)
\left(E_0 \left(q^{[j]}\right) 
    + V_J \left(q^{[j]}\right) 
    - E_0 \left(q^{[i]}\right)
    - V_J \left(q^{[i]}\right)\right)~, 
\label{Eqn35}
\end{equation}
for $T$-exchange, and
\begin{equation}
\Delta = \beta_I 
\left(V_J\left(q^{[j]}\right) - 
      V_J\left(q^{[i]}\right) -
      V_{J+1}\left(q^{[j]}\right) + 
      V_{J+1}\left(q^{[i]}\right)\right)~,
\label{Eqn36}
\end{equation}
for $\lambda$-exchange.
By this procedure, the random walk 
in the reaction coordinate space as well as in the temperature
space can be realized.

\subsection{From Multidimensional REM to Multidimensional MUCA and ST}
The formulations of MREM give multidimensional/multivariable
extensions of REMUCA and REST \cite{RevO2}.
In REMUCA and in REST, the multicanonical weight factor 
and the simulated tempering weight factor are determined from
the results of a short REM simulation, respectively.
The results of a short MREM simulation can therefore be used to 
determine the weight factors
for multidimensional/multivariable MUCA and ST simulatoins, where
random walks in multidimensional ``energy'' and ``parameter''
space are realized \cite{RevO2}.
Here, we give more details.

We consider a simple example with the following potential
energy:
\begin{equation}
E_{\lambda} (q) = E_0 (q) + \lambda V(q)~.
\label{Eqn40}
\end{equation}
In the two-dimensional multicanonical ensemble
each state is weighted by the multicanonical
weight factor $W_{\rm mu}(E_0,V)$ so that a uniform potential energy
distribution both in $E_0$ and $V$ may be obtained:
\begin{equation}
 P_{\rm mu}(E_0,V) \propto n(E_0,V) W_{\rm mu}(E_0,V) \equiv {\rm const}~,
\label{Eqn41}
\end{equation}
where $n(E_0,V)$ is the two-dimensional density of states.
This implies that
\begin{equation}
W_{\rm mu}(E_0,V) \equiv \exp \left[-\beta_0 E_{\rm mu}(E_0,V;T_0) \right]
= \frac{1}{n(E_0,V)}~,
\label{Eqn42}
\end{equation}
where we have chosen an arbitrary reference
temperature, $T_0 = 1/k_{\rm B} \beta_0$, and
the ``{\it multicanonical potential energy}''
is defined by
\begin{equation}
 E_{\rm mu}(E_0,V;T_0) \equiv k_{\rm B} T_0 \ln n(E_0,V)~.
\label{Eqn43}
\end{equation}
The two-dimensional MUCA MC simulation can be performed with the
following transition probability from state $x$ with potential
energy $E_0 + \lambda V$ to state $x^{\prime}$ with
potential energy
${E_0}^{\prime} + \lambda V^{\prime}$ 
(see Eq.~(\ref{eqn8})):
\begin{equation}
w(x \rightarrow x^{\prime})
= {\rm min} \left(1,\frac{W_{\rm mu}({E_0}^{\prime},V^{\prime})}
{W_{\rm mu}(E_0,V)}\right)
= {\rm min} \left(1,\frac{n(E_0,V)}{n({E_0}^{\prime},V^{\prime})}\right)~.
\label{Eqn44}
\end{equation}
The MD algorithm in the two-dimensional multicanonical ensemble
also naturally follows from Eq.~(\ref{eqn6}), in which the
regular constant temperature MD simulation
(with $T=T_0$) is performed by replacing $E$ by $E_{\rm mu}$
in Eq.~(\ref{eqn4f2}) (see Eq.~(\ref{eqn9a})):  
\begin{equation}
\dot{\bs{p}}_k ~=~ - \frac{\partial E_{\rm mu}(E_0,V;T_0)}{\partial \bs{q}_k}
- \frac{\dot{s}}{s}~\bs{p}_k~.
\label{Eqn45}
\end{equation}

In the two-dimensional simulated tempering, the parameter
set $(T, \lambda)$ become dynamical variables, 
and both the configuration and the parameter set are updated
during the simulation with a weight (see Eq.~(\ref{Eqn1})):
\begin{equation}
W_{\rm ST} (E_{\lambda};T,\lambda) 
= \exp \left(-\beta E_{\lambda} + f(T,\lambda) \right)~,
\label{Eqn46}
\end{equation}
where the function $f(T,\lambda)$ is chosen so that 
the probability distribution
of the two-dimensional parameter set is flat (see Eq.~(\ref{Eqn2})):
\begin{equation}
\begin{array}{rl}
P_{\rm ST}(T,\lambda) &= \int dE_0 dV~ n(E_0,V)~ 
W_{{\rm ST}} (E_{\lambda};T,\lambda) \\
&= \int dE_0 dV~ n(E_0,V)~ 
\exp \left(-\beta E_{\lambda} + f(T,\lambda) \right)
= {\rm const}~.
\end{array}
\lab{Eqn47}
\end{equation}

In the numerical work we discretize the parameter
set in $M = N_T \times L$ different values, 
($T_I,\lambda_J)$ ($I=1, \cdots, N_T, J=1, \cdots, L$).  
Without loss of
generality we can order the parameters
so that $T_1 < T_2 < \cdots < T_{N_T}$ and
$\lambda_1 < \lambda_2 < \cdots < \lambda_L$.  
The free energy $f$ is now written as $f_{I,J}=f(T_I,\lambda_J)$.
Once the initial configuration and
the initial parameter set are chosen,
the two-dimensional ST is
then realized by alternately
performing the following two steps:
\begin{enumerate}
\item A canonical MC or MD simulation at the fixed parameter
set $(T_I,\lambda_J)$ is carried out for a certain steps.
\item One of the parameters in the parameter set $(T_I,\lambda_J)$ 
is updated to the neighboring values with the configuration and
the other parameter fixed.  The transition probability of
this parameter-updating
process is given by the following Metropolis criterion:
\begin{equation}
w(T_I \rightarrow T_{I \pm 1})
= {\rm min}\left(1,\frac{W_{\rm ST}(E_{\lambda_J};T_{I \pm 1},\lambda_J)}
{W_{\rm ST}(E_{\lambda_J};T_I,\lambda_J)}\right)
= {\rm min}\left(1,\exp \left( - \Delta \right)\right)~,
\label{Eqn48}
\end{equation}
where
\begin{equation}
\Delta = \left(\beta_{I \pm 1} - \beta_I \right) E_{\lambda_J}
- \left(f_{I \pm 1,J} - f_{I,J} \right)~,
\label{Eqn49}
\end{equation}
for $T$-update, and 
\begin{equation}
w(\lambda_J \rightarrow \lambda_{J \pm 1})
= {\rm min}\left(1,\frac{W_{\rm ST}(E_{\lambda_{J \pm 1}};T_I,\lambda_{J \pm 1})}
{W_{\rm ST}(E_{\lambda_J};T_I,\lambda_J)}\right)
= {\rm min}\left(1,\exp \left( - \Delta \right)\right)~,
\label{Eqn50}
\end{equation}
where
\begin{equation}
\begin{array}{rl}
\Delta &= \beta_I \left( E_{\lambda_{J \pm 1}} - E_{\lambda_J} \right)
- \left(f_{I,J \pm 1} - f_{I,J} \right) \\
&= \beta_I (\lambda_{J \pm 1} - \lambda_J) V
- \left(f_{I,J \pm 1} - f_{I,J} \right)~,
\end{array}
\label{Eqn51}
\end{equation}
for $\lambda$-update.
\end{enumerate}

Finally, we present the corresponding MREM.
We prepare $N_T$ temperatures and $L$ $\lambda$ parameters,
which makes the total
number of replicas $M=N_T \times L$.
We perform the following replica-exchange processes alternately:
\begin{enumerate}
\item Exchange pairs of replicas corresponding to neighboring temperatures,
$T_I$ and $T_{I+1}$ 
(We refer to this process as $T$-exchange.)
\item Exchange pairs of replicas corresponding to 
``neighboring'' $\lambda$ parameters,
$\lambda_J$ and $\lambda_{J+1}$ 
(We refer to this process as $\lambda$-exchange.)
\end{enumerate}
The acceptance criterion for these 
replica exchanges is given by 
Eq.~(\ref{eq15}), where Eq.~(\ref{eqn21}) now reads 
\begin{equation}
\Delta = \left(\beta_{I} - \beta_{I+1} \right)
\left(E_{\lambda_J} \left(q^{[j]}\right) 
    - E_{\lambda_J} \left(q^{[i]}\right)\right)~,
\label{Eqn52}
\end{equation}
for $T$-exchange, and
\begin{equation}
\begin{array}{rl}
\Delta &= \beta_I 
\left(E_{\lambda_J}\left(q^{[j]}\right) - 
      E_{\lambda_J}\left(q^{[i]}\right) -
      E_{\lambda_{J+1}}\left(q^{[j]}\right) + 
      E_{\lambda_{J+1}}\left(q^{[i]}\right)\right) \\
&= \beta_I 
\left(\lambda_J - \lambda_{J+1}\right)
\left(V\left(q^{[j]}\right) - 
      V\left(q^{[i]}\right)\right)~,
\end{array}
\label{Eqn53}
\end{equation}
for $\lambda$-exchange.

After a short MREM simulation, we can use the
multiple-histogram reweighting
techniques to obtain $n(E_0,V)$ and $f_{I,J}$.  
Let $N_{I,J}(E_0,V)$ and $n_{I,J}$
be respectively 
the potential-energy histogram and the total number of
samples obtained for the parameter set
$(T_I,\lambda_J)$.
The WHAM equations 
are then given by 
\begin{equation}
n(E_0,V)
= \frac{\dis{\sum_{I=1}^{N_T} \sum_{J=1}^L N_{I,J}(E_0,V)}} 
{\dis{\sum_{I=1}^{N_T} \sum_{J=1}^L 
n_{I,J}~\exp \left(f_{I,J}-\beta_I 
E_{\lambda_J}\right)}}~,
\label{Eqn54}
\end{equation}
and for each 
$I$ and $J$ ($I=1, \cdots, N_T, J=1, \cdots, L$)  
\begin{equation}
\exp \left( -f_{I,J} \right) = \sum_{E_0,V} n(E_0,V) \exp 
\left(-\beta_I E_{\lambda_J}\right)~.
\label{Eqn55}
\end{equation}
These equations are solved self-consistently by iteration
for $n(E_0,V)$ and $f_{I,J}$.

Hence, we can determine
the multidimensional multicanonical weight factor $W_{\rm mu}(E_0,V)$
and the multidimensional simulated tempering weight factor
$W_{\rm ST}(E_{\lambda_J};T_I,\lambda_J)$.  The former is given by
\begin{equation}
W_{\rm mu}(E_0,V) = \frac{1}{n(E_0,V)}~,
\label{Eqn56}
\end{equation}
and the latter is given by 
\begin{equation}
W_{\rm ST} \left( E_{\lambda_J};T_I,\lambda_J \right)
= \exp \left(-\beta_I E_{\lambda_J} + f_{I,J} \right)~.
\label{Eqn57}
\end{equation}

\section{SIMULATION RESULTS}
We first compare the performances of REM, MUCAREM, and REMUCA.
The accuracy of average quantities calculated depend on the ``quality'' of the 
random walk in the potential energy space, and the measure for this quality 
can be given by the number of tunneling events \cite{MUCA2,MSO03b}.  One 
tunneling event is defined by a trajectory that goes from 
$E_{\rm H}$ to $E_{\rm L}$ and back, where $E_{\rm H}$ and $E_{\rm L}$ are the 
values near the highest energy and the lowest energy, respectively, which the 
random walk can reach. If $E_{\rm H}$ is sufficiently high, the trajectory 
gets completely uncorrelated when it reaches $E_{\rm H}$.  On the other hand, 
when the trajectory reaches near $E_{\rm L}$, it tends to get trapped in 
local-minimum states.  We thus consider that the more tunneling events we 
observe during a fixed number of MC/MD steps, the more efficient the method is 
as a generalized-ensemble algorithm (or, the average quantities obtained by 
the reweighting techniques are more reliable).  

The first example is Monte Carlo simulations of the system of a 17-residue 
fragment of ribonuclease T1 in implicit solvent (expressed by the solvent 
accessible surface area) \cite{MSO03b}.  The amino-acid sequence is 
Ser-Ser-Asp-Val-Ser-Thr-Ala-Gln-Ile-Ala-Ala-Tyr-Lys-Leu-His-Glu-Asp.  
The energy function $E_{\rm TOT}$ that we used is the sum of the 
conformational energy term of the solute $E_{\rm P}$ and the solvation 
free energy term $E_{\rm SOL}$ for the interaction of the peptide with 
the surrounding solvent: $E_{\rm TOT}=E_{\rm P} + E_{\rm SOL}$.  Here, the 
solvation term $E_{\rm SOL}$ is given by the sum of the terms that are 
proportional to the solvent-accessible surface area of the atomic groups 
of the solute.  The parameters in the conformational energy as well as the 
molecular geometry were taken from ECEPP/2.  The parameters of the solvent term were 
adopted from Ref.~\cite{SOL1}.  The computer code KONF90 \cite{KONF1,KONF2} was used, 
and MC simulations based on the REM, MUCAREM, and REMUCA were performed.  For the 
calculation of a solvent-accessible surface area, we used the computer code 
NSOL \cite{SOL2}.  The dihedral angles $\phi$ and $\psi$ in the main chain and $\chi$ 
in the side chain constituted the variables to be updated in the MC 
simulations.  The number of degrees of freedom for the peptide is 80.  One 
MC sweep consists of updating all these angles once with Metropolis 
evaluation for each update.  The simulations were started from randomly 
generated conformations.  In Table~\ref{tab:1} we list the number of 
tunneling events in REM, MUCAREM, and REMUCA simulations of the same system \cite{MSO03b}.

\begin{table}[th]
\begin{center}
\caption{Number of tunneling events in the MC simulations of a fragment of ribonuclease T1 for REM, MUCAREM, and REMUCA simulations}
\label{tab:1}       
%
%
\begin{tabular}{cccc}
\hline\noalign{\smallskip}
Total MC sweeps~ & ~REM~ & ~MUCAREM~ & ~REMUCA~ \\
\noalign{\smallskip}
\hline\noalign{\smallskip}
$2 \times 10^6$ & 2 & 9 & 18 \\
$3 \times 10^6$ & 5 & 16 & 29 \\
$4 \times 10^6$ & 9 & 22 & 38 \\
\noalign{\smallskip}\hline
\end{tabular}
\end{center}
\end{table}
\begin{table}[bh]
\begin{center}
\caption{Number of tunneling events in the MD simulations of three peptides in expicit water for REM, MUCAREM, and REMUCA simulations}
\label{tab:2}       
%
%
\begin{tabular}{cccccc}
\hline\noalign{\smallskip}
Peptide & No. of atoms & ~Total MD steps~ & REMD~ & MUCAREM~ & REMUCA \\
\noalign{\smallskip}
\hline\noalign{\smallskip}
Alanine dipeptide & 418 & $4 \times 10^6$ & 11 & 40 & 59 \\
Alanine trimer & 876 & $5 \times 10^6$ & 1 & 20 & 29 \\
Met-enkephalin & 1662 & $8 \times 10^6$ & 0 & 12 & 27 \\
\noalign{\smallskip}\hline
\end{tabular}
\end{center}
\end{table}
   
Hence, REMUCA is the most efficient, then MUCAREM, and finally REM.

The next systems are small peptides in explicit water \cite{SO4}.  When we 
consider
explicit water molecules, the problem becomes order-of-magnitude more
difficult than the case with implicit water models.  They are alanine 
dipeptide with 132 water molecules, alanine trimer with 278 water molecules, 
and Met-enkephalin with 526 water molecules.  The force-field, or the potential
energy, that we used 
is AMBER parm96 \cite{AMBER96} for the peptides and TIP3P \cite{TIP3P}
for water molecules.  The peptides 
were placed inside the spheres of water molecules and the harmonic 
constraining forces were imposed in order to prevent the water molecules 
from flying apart. The unit time step, $\Delta t$, was set to 0.5 fsec.
The modified version \cite{SK98,kitao98} of the 
software PRESTO version 2 \cite{PRESTO} was used.
In Table~\ref{tab:2} we list the number of tunneling events in these systems.

\begin{figure}[b]
\centering
\includegraphics[height=8cm]{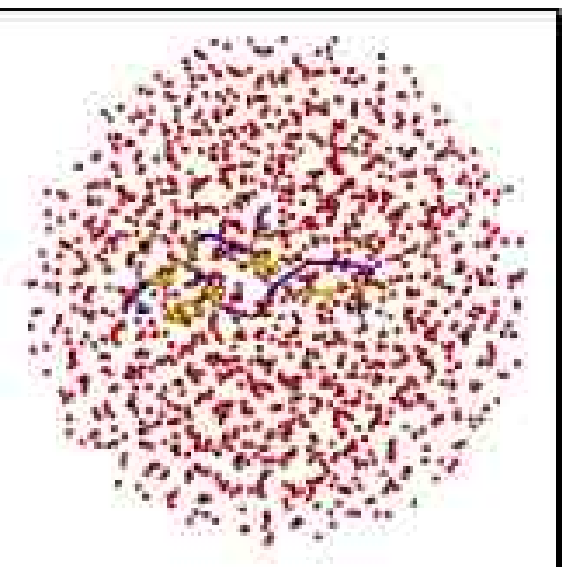}
%
%
\caption{The initial configuration of C-peptide in explicit water, which was 
used in all of the 32 replicas of the first REMD simulation (REMD1 in 
Table~\ref{tab:3}).  The red filled circles stand for the oxygen atoms of 
water molecules.  The number of water molecules is 1387, and they are placed 
in a sphere of radius 22 \AA.  As for the peptide, besides the backbone 
structure (in blue), side chains of only Glu$^-$-2, Phe-8, Arg$^+$-10, 
and His$^+$-12 are shown (in yellow).  The figure was created with 
Molscript \cite{Molscript} and Raster3D \cite{Raster3D}.}
\label{fig:1}       
\end{figure}

The last system is the C-peptide of ribonuclease A in explicit water
\cite{SO5}.
In the model of simulations, the N-terminus and the C-terminus of the 
C-peptide analogue were blocked with the acetyl group and the N-methyl group, 
respectively.  The number of amino acids is 13 and the amino-acid sequence is: 
Ace-Ala-Glu$^-$-Thr-Ala-Ala-Ala-Lys$^+$-Phe-Leu-Arg$^+$-Ala-His$^+$-Ala-Nme 
\cite{shoe87,shoe90}. The initial 
configuration of our simulation was first generated by a high temperature 
molecular dynamics simulation (at $T = 1000$ K) in gas phase, starting from 
a fully extended conformation. We randomly selected one of the structures 
that do not have any secondary structures such as $\alpha$-helix and 
$\beta$-sheet. The peptide was then solvated in a sphere of radius 22 \AA, 
in which 1387 water molecules were included (see Fig. 1).  Harmonic restraint 
was applied to prevent the water molecules from going out of the sphere.  
The total number of atoms is 4365.  The dielectric constant was set equal 
to 1.0.
The force-field parameters for protein were taken from the all-atom version 
of AMBER parm99 \cite{AMBER99}, which was 
found to be suitable for studying helical peptides \cite{yso04}, 
and TIP3P model \cite{TIP3P} was used for water molecules.
The unit time step, $\Delta t$, was set to 0.5 fsec.

In Table~\ref{tab:3} the essential parameters in the simulations performed 
in this article are summarized.  
\begin{table}[b]
\begin{center}
\caption{Summary of parameters in REMD, MUCAREM, and REMUCA simulations}
\label{tab:3}       
%
%
\begin{tabular}{lcll}
\hline\noalign{\smallskip}
 & ~~Number of~~ & ~~~Temperature, & ~~MD steps per  \\
 & ~~replicas, $M$~~ & $~~~T_m$ (K) ($m=1, \cdots, M$) & ~~replica \\
\noalign{\smallskip}
\hline\noalign{\smallskip}
REMD1* & 32 & 250, 258, 267, 276, 286, 295, 305, & ~~~$2.0 \times 10^5$ \\
      &    & 315, 326, 337, 348, 360, 372, 385, & \\
     &    & 398, 411, 425, 440, 455, 470, 486, & \\
     &    & 502, 519, 537, 555, 574, 593, 613, & \\
     &    & 634, 655, 677, 700 & \\
MUCAREM1 & 4 & 360, 440, 555, 700 & $~~~2.0 \times 10^6$ \\
REMUCA1 & 1 & 700 & ~~~$3.0 \times 10^7$ \\
\noalign{\smallskip}\hline
\end{tabular}
\end{center}
* REMD1 stands for the replica-exchange molecular dynamics simulation, 
MUCAREM1 stands for the multicanonical replica-exchange molecular dynamics 
simulation, and REMUCA1 stands for the final multicanonical molecular 
dynamics simulation (the production run) of REMUCA.  The results of REMD1 
were used to determine the multicanonical weight factors for MUCAREM1, and 
those of MUCAREM1 were used to determine the multicanonical weight factor 
for REMUCA1.
\end{table}

We first performed a REMD simulation with 32 replicas for 100 psec per 
replica (REMD1 in Table~\ref{tab:3}).  During this REMD simulation, replica exchange 
was tried every 200 MD steps. Using the obtained potential-energy histogram 
of each replica as input data to the multiple-histogram analysis in 
Eqs. (4) and (5), we obtained the first estimate of the multicanonical 
weight factor, or the density of states. We divided this multicanonical 
weight factor into four multicanonical weight factors that cover different 
energy regions \cite{SO3,MSO03,MSO03b}
and assigned these multicanonical weight 
factors into four replicas (the weight factors cover the potential energy 
ranges from $-13791.5$ to $-11900.5$ kcal/mol, from $-12962.5$ to 
$-10796.5$ kcal/mol, from $-11900.5$ to $-9524.5$ kcal/mol, and from 
$-10796.5$ to $-8293.5$ kcal/mol).  We then carried out a MUCAREM 
simulation with four replicas for 1 nsec per replica (MUCAREM1 in 
Table~\ref{tab:3}), 
in which replica exchange was tried every 1000 MD steps. We again used the 
potential-energy histogram of each replica as the input data to the 
multiple-histogram analysis and finally obtained the multicanonical weight 
factor with high precision. As a production run, we carried out a 15 nsec 
multicanonical MD simulation with one replica (REMUCA1 in 
Table~\ref{tab:3}) and the 
results of this production run were analyzed in detail.

\begin{figure}[ht]
\centering
\includegraphics[height=13cm]{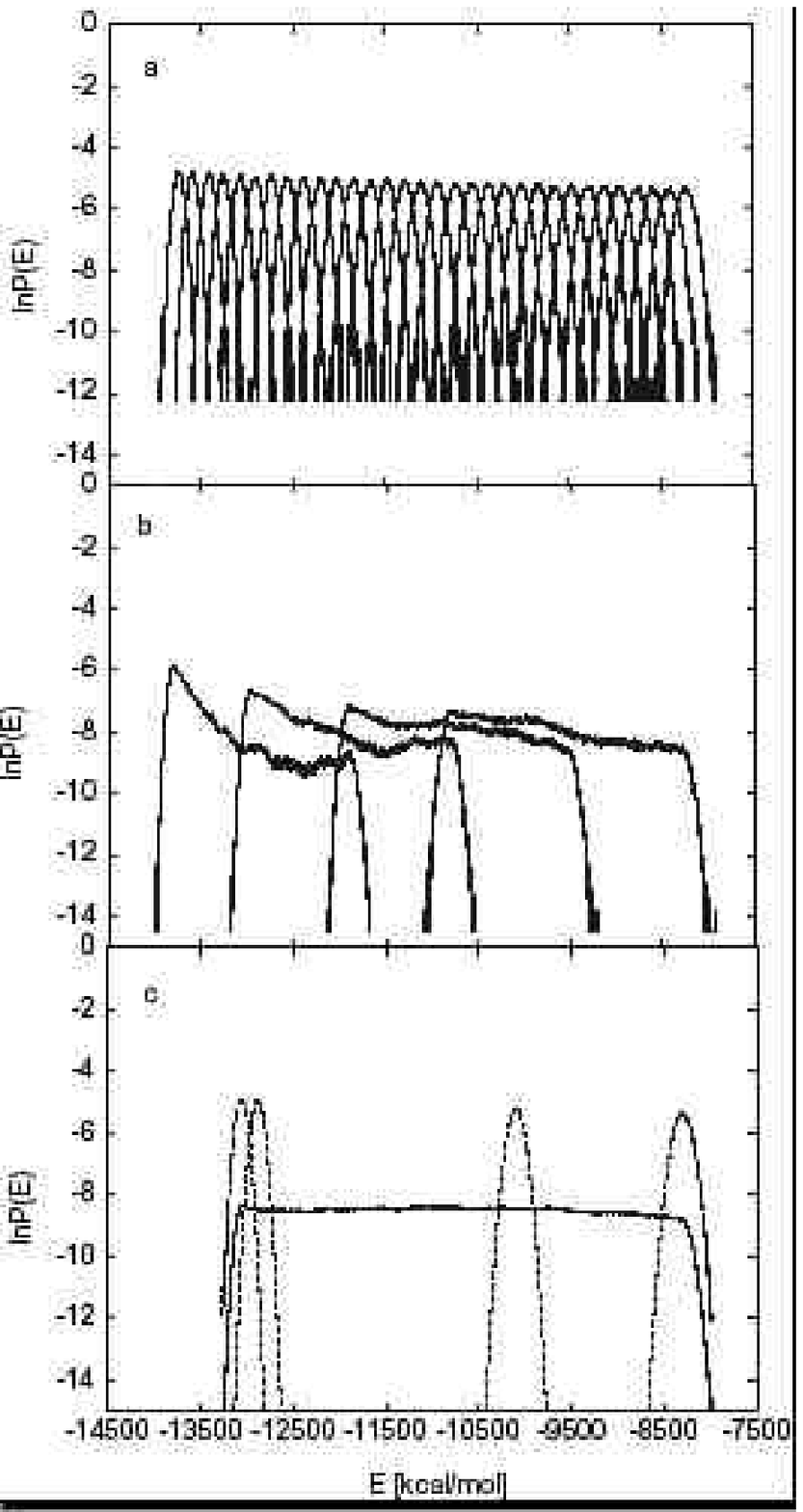}
%
%
\caption{Probability distributions of potential energy of the C-peptide 
system obtained from (a) REMD1, (b) MUCAREM1, and (c) REMUCA1.  See 
Table~\ref{tab:3} for the parameters of the simulations.   Dashed curves 
in (c) are the reweighted canonical distributions at 290, 300, 500, and 
700 K (from left to right).}
\label{fig:2}       
\end{figure}
In Fig.~\ref{fig:2} we show the probability distributions of potential 
energy that were obtained from the above three generalized-ensemble 
simulations, namely, REMD1, MUCAREM1, and REMUCA1.  We see in 
Fig.~\ref{fig:2}(a) that there are enough overlaps between all pairs of 
neighboring canonical distributions, suggesting that there were sufficient 
numbers of replica exchange in REMD1.  We see in Fig.~\ref{fig:2}(b) that 
there are good overlaps between all pairs of neighboring multicanonical 
distributions, implying that MUCAREM1 also performed properly.  Finally, 
the multicanonical distribution in Fig.~\ref{fig:2}(c) is completely flat 
between around $-13000$ kcal/mol and around $-8000$ kcal/mol.  The results 
suggest that a free random walk was realized in this energy range.

\begin{figure}
\centering
\includegraphics[height=4.8cm]{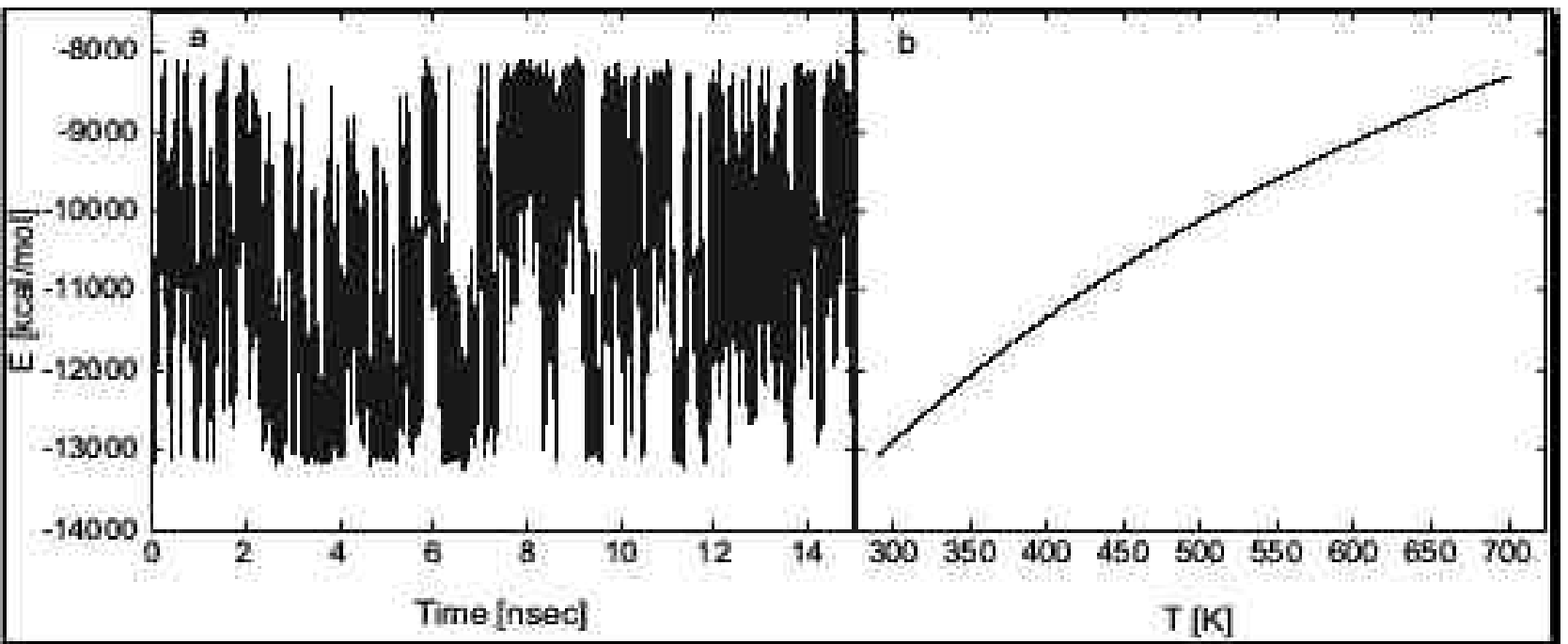}
%
%
\caption{Time series of potential energy of the C-peptide system from the 
REMUCA production run (REMUCA1 in Table~\ref{tab:3}) (a) and the average 
potential energy as a function of temperature (b).  The latter was obtained 
from the trajectory of REMUCA1 by the single-histogram reweighting techniques.}
\label{fig:3}       
\end{figure}

In Fig.~\ref{fig:3}a we show the time series of potential energy from REMUCA1.
We indeed observe a random walk covering as much as 5000 kcal/mol of energy 
range (note that 23 kcal/mol $\approx$ 1 eV).  We show in Fig.~\ref{fig:3}(b) 
the average potential energy as a function of temperature, which was obtained 
from the trajectory of REMUCA1 by the reweighting techniques.  The average 
potential energy monotonically increases as the temperature increases.

Here, we took $E_{\rm H} = -8250$ kcal/mol and $E_{\rm L} = -12850$ kcal/mol 
for the 
measurement of the tunneling events.  The random walk in REMUCA1 yielded as 
many as 55 tunneling events in 15 nsec.  The corresponding numbers of 
tunneling events for REMD1 and for MUCAREM1 were 0 in 3.2 nsec and 5 in 4 nsec,
respectively.  Hence, REMUCA is the most efficient and reliable among the 
three generalized-ensemble algorithms.

\begin{figure}
\centering
\includegraphics[width=9cm]{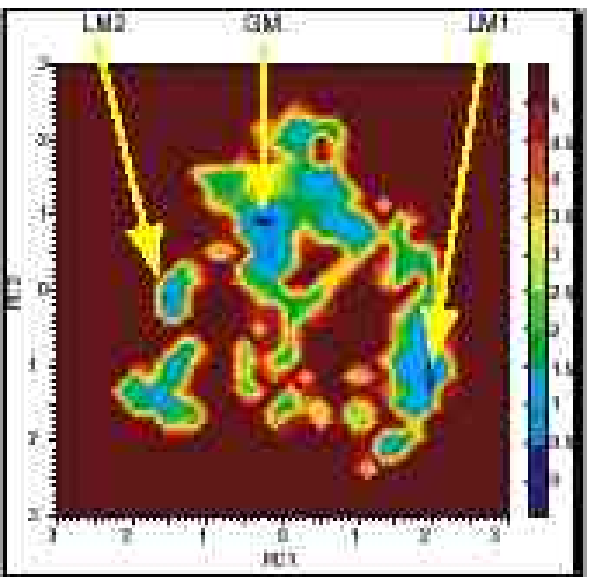}
%
%
\caption{Potential of mean force (kcal/mol) of the C-peptide system along the 
first two principal components at 300 K. The free energy was calculated from 
the results of REMUCA production run (REMUCA1 in Table~\ref{tab:3}) by the 
single-histogram reweighting techniques and normalized so that the 
global-minimum state (GM) has the value zero.  GM, LM1, and LM2 represent 
three distinct minimum free-energy states.}
\label{fig:5}       
\end{figure}

\begin{figure}
\centering
\includegraphics[height=4.5cm]{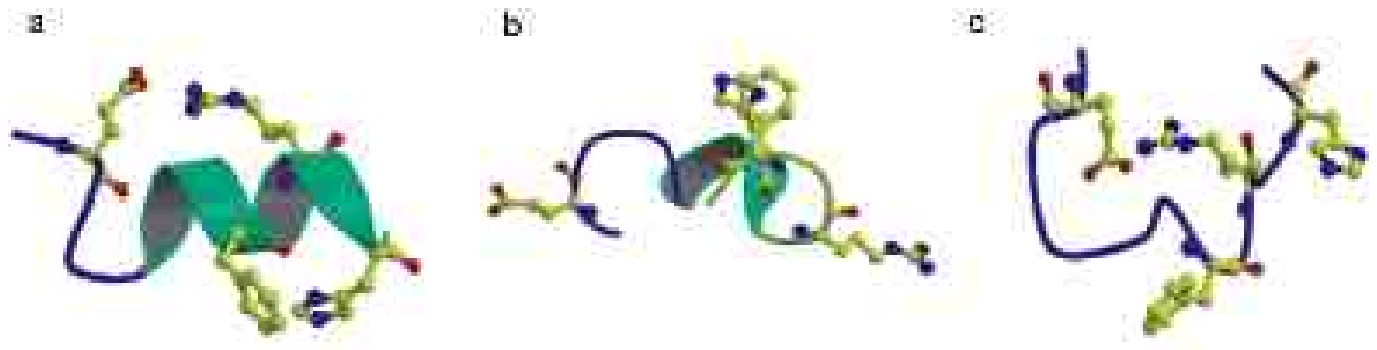}
%
%
\caption{The representative structures at the global-minimum free-energy state 
((a) GM) and the two local-minimum states ((b) LM1 and (c) LM2).  As for the 
peptide structures, besides the backbone structure, side chains of only 
Glu$^-$-2, Phe-8, Arg$^+$-10, and His$^+$-12 are shown in ball-and-stick model.}
\label{fig:6}       
\end{figure}

In Fig.~\ref{fig:5} the potential of mean force (PMF), or free energy, 
along the first two principal component axes at 300 K is shown.  There exist 
three distinct minima in the free-energy landscape, which correspond to three 
local-minimum-energy states. We show representative conformations at these 
minima in Fig.~\ref{fig:6}.  The structure of the global-minimum free-energy 
state (GM) has a partially distorted $\alpha$-helix with the salt bridge 
between Glu$^-$-2 and Arg$^+$-10. The structure is in good agreement with the 
experimental structure obtained by both NMR and X-ray experiments.  In this 
structure there also exists a contact between Phe-8 and His$^+$-12. This 
contact is again observed in the corresponding residues of the X-ray structure.
At LM1 the structure has a contact between Phe-8 and His$^+$-12, but the salt 
bridge between Glu$^-$-2 and Arg$^+$-10 is not formed.  On the other hand, the 
structure at LM2 has this salt bridge, but it does not have a contact between 
Phe-8 and His$^+$-12. Thus, only the structures at GM satisfy all of the 
interactions that have been observed by the X-ray and other experimental 
studies.

Finally, we remark that the largest peptide in explicit water that
we have succeeded in folding into the native structure from random
initial conformations is so far the 16-residue C-terminal $\beta$-hairpin
of streptococcal protein G B1 domain, which was accomplished by 
MUCAREM simulations with eight replicas \cite{yso07}.

\section{CONCLUSIONS}

In this article we have reviewed some of powerful
generalized-ensemble
algorithms for both Monte Carlo simulations and molecular dynamics
simulations.
A simulation in generalized ensemble realizes a random
walk in potential energy space, alleviating the
multiple-minima problem that is a common difficulty
in simulations of complex systems with many degrees
of freedom.

Detailed formulations of the three well-known
generalized-ensemble algorithms, namely,
multicanonical algorithm (MUCA),
simulated tempering (ST), and replica-exchange method (REM), 
were given.  

We then introduced several new generalized-ensemble
algorithms that combine the merits of the above three methods. 

The question is then which method is the most recommended.
Our criterion for the
effectiveness of generalized-ensemble algorithms was 
how many random walk cycles (tunneling events)
in potential energy space
between the high-energy region and low-energy region 
are realized within a fixed 
number of total MC (or MD) steps.
We found that once the optimal MUCA weight factor is
obtained, MUCA (and REMUCA) is the most effective
(i.e., has the most number of
tunneling events), and REM is the least \cite{MSO03}.
We also found that once the optimal ST weight factor
is obtained, ST (and REST) has more tunneling events
than REM \cite{MO4,STREM}.  Moreover, we compared
the efficiency of  
Berg's recursion \cite{MUCAW},
Wang-Landau method \cite{Landau1,Landau2},
and REMUCA/MUCAREM as methods for the multicanonical weight factor
determination in two-dimensional 10-state Potts model
and found that the three methods are about equal in efficiency
\cite{NSMO}--\cite{O03}.

Hence, the answer to the above question will depend on 
how much time one is willing to (or forced to)
spend in order to determine the MUCA or ST weight factors.
Given a problem, the first choice is REM because of its
simplicity (no weight factor determination is required).
If REM turns out to be insufficient or too much 
time-consuming (like the case with
first-order phase transitions), then other
more powerful algorithms such as MUCAREM and STREM
are recommended.

\vspace{0.5cm}
\noindent
{\bf Acknowledgements}: \\
This work was supported, in part, by Grants-in-Aid
for Scientific Research in Priority Areas (``Water and Biomolecules''
and ``Membrane Interface'')
and for the Next Generation Super Computing Project, Nanoscience Program
from the Ministry of
Education, Culture, Sports, Science and Technology (MEXT), Japan.\\


%
%
%
%


\noindent
\begin{thebibliography}{(000)}


\bibitem{RevHO1} U.H.E. Hansmann, Y. Okamoto:  Generalized-ensemble
approach for protein folding simulations.
In: {\it Annual Reviews of Computational Physics VI}, ed by D. Stauffer
(World Scientific, Singapore, 1999) pp. 129--157.
\bibitem{RevMSO} A. Mitsutake, Y. Sugita, Y. Okamoto: 
{\it Biopolymers (Peptide Science)} \textbf{60}, 96--123 (2001).
\bibitem{RevSO} Y. Sugita, Y. Okamoto: 
Free-energy calculations in protein folding by generalized-ensemble algorithms.
In: {\it Lecture Notes in Computational Science and Engineering}, 
ed by T. Schlick, H.H. Gan
(Springer-Verlag, Berlin, 2002) pp. 304--332; e-print: cond-mat/0102296.

\bibitem{Be02} B.A. Berg:
Comp. Phys. Commun. \textbf{147}, 
52--57 (2002).
\bibitem{RevO2} Y. Okamoto: J. Mol. Graphics Mod.
        \textbf{22}, 425--439 (2004); e-print: cond-mat/0308360.
\bibitem{RevKO} H. Kokubo, Y. Okamoto: Mol. Sim.
        \textbf{32}, 791--801 (2006).
\bibitem{RevIOO} S.G. Itoh, H. Okumura, Y. Okamoto: Mol. Sim.
        \textbf{33}, 47--56 (2007).


\bibitem{FS1} A.M. Ferrenberg, R.H. Swendsen: 
Phys. Rev. Lett. \textbf{61}, 2635--2638 (1988); {\it ibid.}
\textbf{63}, 1658 (1989).
\bibitem{FS2} A.M. Ferrenberg, R.H. Swendsen: 
Phys.  Rev.  Lett. \textbf{63}, 1195--1198 (1989).
\bibitem{WHAM} S. Kumar, D. Bouzida, R.H. Swendsen, P.A. Kollman, 
J.M. Rosenberg: J. Comput. Chem. \textbf{13}, 1011--1021 (1992).
  
\bibitem{SW87} R.H. Swendsen, J.S. Wang:
Phys.  Rev.  Lett. \textbf{58}, 86--88 (1987).

\bibitem{Wolff89} U. Wolff: Phys. Rev. Lett. 
\textbf{62}, 361--364 (1989).
\bibitem{ELM93} H.G. Evertz, G. Lana, M. Marcu: 
Phys. Rev. Lett. \textbf{70}, 875--879 (1993).



\bibitem{MUCA1} B.A. Berg, T. Neuhaus: Phys. Lett. 
\textbf{B267}, 249--253 (1991).
\bibitem{MUCA2} B.A. Berg, T. Neuhaus, Phys. Rev. Lett. 
\textbf{68}, 9--12 (1992).
\bibitem{MUCArev} B.A. Berg: 
{\it Fields Institute Communications} {\bf 26}, 1--24 (2000); also see
e-print: cond-mat/9909236.

\bibitem{RevJanke} W. Janke: 
Physica A \textbf{254}, 164--178 (1998).


\bibitem{Lee} J. Lee: 
Phys. Rev. Lett. \textbf{71}, 211--214 (1993);
ibid. \textbf{71}, 2353 (1993).
\bibitem{MZ} M. Mezei: 
J. Comput. Phys. \textbf{68}, 237--248 (1987).
\bibitem{BK} C. Bartels, M. Karplus: 
 J. Phys. Chem. B \textbf{102}, 865--880 (1998).

\bibitem{US} G.M. Torrie, J.P. Valleau: 
J. Comput. Phys. \textbf{23}, 187--199 (1977).
 
\bibitem{WS02} J.S. Wang, R.H. Swendsen: 
J. Stat. Phys. {\bf 106}, 245--285 (2002).
   
\bibitem{Landau1} F. Wang, D.P. Landau: 
Phys. Rev. Lett. \textbf{86}, 2050--2053 (2001).
\bibitem{Landau2} F. Wang, D.P. Landau: 
Phys. Rev. E \textbf{64}, 056101 (2001).
   
\bibitem{dePablo} Q. Yan, R. Faller, J.J. de Pablo: 
J. Chem. Phys. \textbf{116}, 8745--8749 (2002).

\bibitem{THT04} S. Trebst, D.A. Huse, M. Troyer:
Phys. Rev. E \textbf{70} 046701 (2004).
   
\bibitem{MUCA3} B.A. Berg, T. Celik: Phys. Rev. Lett. 
\textbf{69}, 2292--2295 (1992).
\bibitem{BeHaNe93} B.A. Berg, U.H.E. Hansmann, T. Neuhaus: 
Phys. Rev. B \textbf{47}, 497--500 (1993).
\bibitem{JK95} W. Janke, S. Kappler: Phys. Rev. Lett. 
\textbf{74}, 212--215 (1995).
{\bf 74}, 
\bibitem{MUCA5} B.A. Berg, W. Janke:  Phys. Rev. Lett. 
\textbf{80}, 4771--4774 (1998).
\bibitem{MUCA6} N. Hatano, J.E. Gubernatis: 
Prog. Theor. Phys. (Suppl.) \textbf{138}, 442--447 (2000). 
\bibitem{BBJ00} B.A. Berg, A. Billoire, W. Janke: 
Phys. Rev. B \textbf{61}, 12143--12150 (2000).

\bibitem{HO} U.H.E. Hansmann, Y. Okamoto: 
J. Comput. Chem. \textbf{14}, 1333--1338 (1993).
\bibitem{HO94} U.H.E. Hansmann, Y. Okamoto: 
 Physica {\bf A212}, 415--437 (1994).
\bibitem{HSch} M.H. Hao, H.A. Scheraga: 
        J. Phys. Chem. \textbf{98}, 4940--4948 (1994).
\bibitem{OH} Y. Okamoto, U.H.E. Hansmann: 
   J. Phys. Chem. \textbf{99}, 11276--11287 (1995).


\bibitem{Wilding95} N.B. Wilding: Phys. Rev. E 
\textbf{52}, 602--611 (1995).

\bibitem{KGS} A. Kolinski, W. Galazka, J. Skolnick: 
{\it Proteins} \textbf{26}, 271--287 (1996).
\bibitem{UT} N. Urakami, M. Takasu: 
J. Phys. Soc. Jpn. \textbf{65}, 2694--2699 (1996).
\bibitem{KPV} S. Kumar, P. Payne, M. V{\' a}squez: 
J. Comput. Chem. \textbf{17}, 1269--1275 (1996).

\bibitem{HOE96} U.H.E. Hansmann, Y. Okamoto, F. Eisenmenger: 
Chem. Phys. Lett. {\bf 259}, 321--330 (1996).

\bibitem{HO96a} U.H.E. Hansmann, Y. Okamoto: Phys. Rev. E 
\textbf{54}, 5863--5865 (1996).

\bibitem{HO96b} U.H.E. Hansmann, Y. Okamoto: 
J. Comput. Chem. \textbf{18}, 920--933 (1997).

\bibitem{NNK} N. Nakajima, H. Nakamura, A. Kidera: 
J. Phys. Chem. B \textbf{101}, 817--824 (1997).
\bibitem{BK2} C. Bartels, M. Karplus: 
 J. Comput. Chem. \textbf{18}, 1450--1462 (1997).

\bibitem{HNSKN} J. Higo, N. Nakajima, H. Shirai, A. Kidera, 
H. Nakamura: 
J. Comput. Chem. \textbf{18}, 2086--2092 (1997).


\bibitem{ICK} Y. Iba, G. Chikenji, M. Kikuchi:
J. Phys. Soc. Jpn.
\textbf{67}, 3327--3330 (1998).

\bibitem{MHO} A. Mitsutake, U.H.E. Hansmann, Y. Okamoto: 
{\it J. Mol. Graphics Mod.} {\bf 16}, 226--238; 262--263 (1998).

\bibitem{HO99} U.H.E. Hansmann, Y. Okamoto: 
J. Phys. Chem. B \textbf{103}, 1595--1604 (1999).
\bibitem{SUYH} H. Shimizu, K. Uehara, K. Yamamoto, and Y. Hiwatari: 
Mol. Sim. \textbf{22}, 285--301 (1999).
\bibitem{ONHN} S. Ono, N. Nakajima, J. Higo, H. Nakamura: 
Chem. Phys. Lett. \textbf{312}, 247--254 (1999).
\bibitem{MO2} A. Mitsutake, Y. Okamoto: 
J. Chem. Phys. \textbf{112}, 10638--10647 (2000).
\bibitem{SKGS00} K. Sayano, H. Kono, M.M. Gromiha, and A. Sarai: 
J. Comput. Chem. \textbf{21}, 954--962 (2000).
\bibitem{YCBM} F. Yasar, T. Celik, B.A. Berg, H. Meirovitch: 
J. Comput. Chem. \textbf{21}, 1251--1261 (2000).
\bibitem{MO3} A. Mitsutake, M. Kinoshita, Y. Okamoto, F. Hirata:
Chem. Phys. Lett. \textbf{329}, 295--303 (2000).
\bibitem{CGO02} M.S. Cheung, A.E. Garcia, J.N. Onuchic: 
Proc. Natl. Acad. Sci. U.S.A. \textbf{99}, 685--690 (2002).

\bibitem{KHN02} N. Kamiya, J. Higo, H. Nakamura:
Protein Sci. \textbf{11}, 2297--2307 (2002).
\bibitem{JPS02} S.W. Jang, Y. Pak, S.M. Shin: 
J. Chem. Phys. \textbf{116}, 4782--4786 (2002).


\bibitem{KFN03} J.G. Kim, Y. Fukunishi, H. Nakamura: 
Phys. Rev. E \textbf{67}, 011105 (2003).

\bibitem{RKP03} N. Rathore, T.A. Knotts, IV, J.J. de Pablo: 
J. Chem. Phys. \textbf{118}, 4285--4290 (2003).

\bibitem{TMK03} T. Terada, Y. Matsuo, A. Kidera: 
J. Chem. Phys. \textbf{118}, 4306--4311 (2003).

\bibitem{BNO03} B.A. Berg, H. Noguchi, Y. Okamoto:
Phys. Rev. E \textbf{68}, 036126 (2003).

\bibitem{BJ03} M. Bachmann, W. Janke:
Phys. Rev. Lett. \textbf{91}, 208105 (2003).

\bibitem{OO03} H. Okumura, Y. Okamoto: 
Chem. Phys. Lett. \textbf{383}, 391--396 (2004).

\bibitem{Muna} T. Munakata, S. Oyama: 
Phys. Rev. E \textbf{54}, 4394--4398 (1996). 


\bibitem{ST1} A.P. Lyubartsev, A.A. Martinovski, S.V. Shevkunov, 
P.N. Vorontsov-Velyaminov: 
J. Chem. Phys. \textbf{96}, 1776--1783 (1992).
\bibitem{ST2} E. Marinari, G. Parisi: 
Europhys. Lett. \textbf{19}, 451--458 (1992).
\bibitem{STrev} E. Marinari, G. Parisi, J.J. Ruiz-Lorenzo: 
In: {\it Spin Glasses and Random Fields}, ed by A.P. Young
(World Scientific, Singapore, 1998) pp. 59--98.
\bibitem{IRB1} A. Irb{\"a}ck, F. Potthast: 
J. Chem. Phys. \textbf{103}, 10298--10305 (1995).
\bibitem{IRB2} A. Irb{\"a}ck, E. Sandelin: 
J. Chem. Phys. \textbf{110}, 12256--12262 (1999).

\bibitem{SmBr} G.R. Smith, A.D. Bruce: 
Phys. Rev. E \textbf{53}, 6530--6543 (1996).
\bibitem{H97c} U.H.E. Hansmann: 
Phys. Rev. E \textbf{56}, 6200--6203 (1997).
\bibitem{MUCAW} B.A. Berg: 
Nucl. Phys. B (Proc. Suppl.) \textbf{63A-C}, 982--984 (1998).
\bibitem{Janke03} W. Janke: 
Histograms and all that.
In: {\it Computer Simulations of Surfaces and Interfaces V}, 
NATO Science Series, II Mathematics, Physics and Chemistry
Vol.~{\bf 114}, Proceedings of the NATO Advanced Study Institute, ed by 
B. D{\" u}nweg, D.P. Landau, A.I. Milchev 
(Kluwer, Dordrecht, 2003), pp. 137--157.


\bibitem{RE1} K. Hukushima, K. Nemoto:
J. Phys. Soc. Jpn. \textbf{65}, 1604--1608 (1996).
\bibitem{RE1b} K. Hukushima, H. Takayama, K. Nemoto:
Int. J. Mod. Phys. C \textbf{7}, 337--344 (1996).

\bibitem{RE2} C.J. Geyer: In: {\it Computing Science and Statistics:
 Proc. 23rd Symp. on the Interface}, ed by E.M. Keramidas
(Interface Foundation, Fairfax Station, 1991) pp. 156--163.
\bibitem{RE3} R.H. Swendsen, J.-S. Wang: 
Phys. Rev. Lett. \textbf{57}, 2607--2609 (1986).
\bibitem{KT} K. Kimura, K. Taki: In: {\it Proc. 13th IMACS World Cong. 
on Computation and Appl. Math. (IMACS '91)}, ed by
R. Vichnevetsky, J.J.H. Miller,
vol. 2, pp. 827--828.
\bibitem{JWK} D.D. Frantz, D.L. Freeman, 
J.D. Doll: 
J. Chem. Phys. \textbf{93}, 2769--2784 (1990).
\bibitem{RE4} M.C. Tesi, E.J.J. van Rensburg, E. Orlandini,
S.G. Whittington: 
J. Stat. Phys. \textbf{82}, 155--181 (1996).
\bibitem{IBArev} Y. Iba: 
Int. J. Mod. Phys. C \textbf{12}, 623--656 (2001).

\bibitem{H97} U.H.E. Hansmann: 
Chem. Phys. Lett. \textbf{281}, 140--150 (1997). 

\bibitem{SO} Y. Sugita, Y. Okamoto: 
Chem. Phys. Lett. \textbf{314}, 141--151 (1999).

\bibitem{WD} M.G. Wu, M.W. Deem:
Mol. Phys. \textbf{97}, 559--580 (1999). 

\bibitem{SKO} Y. Sugita, A. Kitao, Y. Okamoto: 
J. Chem. Phys. \textbf{113}, 6042--6051 (2000).

\bibitem{WEK03} C.J. Woods, J.W. Essex, M.A. King:
J. Phys. Chem. B \textbf{107}, 13703--13710 (2003).
   
\bibitem{SO3} Y. Sugita, Y. Okamoto: 
Chem. Phys. Lett. \textbf{329}, 261--270 (2000).

\bibitem{MO4} A. Mitsutake, Y. Okamoto: 
Chem. Phys. Lett. \textbf{332}, 131--138 (2000).


\bibitem{Kol} D. Gront, A. Kolinski, J. Skolnick: 
J. Chem. Phys. \textbf{113}, 5065--5071 (2000).
\bibitem{Verk01} G.M. Verkhivker, P.A. Rejto, D. Bouzida, 
S. Arthurs, A.B. Colson, S.T. Freer, D.K. Gehlhaar, V. Larson, 
B.A. Luty, T. Marrone, P.W. Rose: 
Chem. Phys. Lett. \textbf{337}, 181--189 (2001).

\bibitem{FWT02} H. Fukunishi, O. Watanabe, S. Takada:
J. Chem. Phys. \textbf{116}, 9058--9067 (2002).

\bibitem{MSO03} A. Mitsutake, Y. Sugita, Y. Okamoto:
J. Chem. Phys. \textbf{118}, 6664--6675 (2003).
\bibitem{MSO03b} A. Mitsutake, Y. Sugita, Y. Okamoto:
J. Chem. Phys. \textbf{118}, 6676--6688 (2003).

\bibitem{SR03} A. Sikorski, P. Romiszowski: 
Biopolymers \textbf{69}, 391--398 (2003).


\bibitem{LHH03} C.Y. Lin, C.K. Hu, U.H.E. Hansmann: 
Proteins \textbf{52}, 436--445 (2003).

\bibitem{LMMO03} G. La Penna, A. Mitsutake, M. Masuya, Y. 
Okamoto: Chem. Phys. Lett. \textbf{380}, 609--619 (2003).


\bibitem{FD} M. Falcioni, M.W. and Deem, M.W. (1999)
J. Chem. Phys. \textbf{110}, 1754--1766. 

\bibitem{YP} Q. Yan, J.J. de Pablo: 
J. Chem. Phys. \textbf{111}, 9509--9516 (1999). 

\bibitem{NOSMO} T. Nishikawa, H. Ohtsuka, Y. Sugita, M. 
Mikami, Y. Okamoto: 
Prog. Theor. Phys. (Suppl.) \textbf{138}, 270--271 (2000).

\bibitem{Yama} R. Yamamoto, W. Kob: 
Phys. Rev. E \textbf{61}, 5473--5476 (2000). 


\bibitem{K02} D.A. Kofke: 
J. Chem. Phys. \textbf{117}, 6911--6914 (2002). 

\bibitem{OKOM} T. Okabe, M. Kawata, Y. Okamoto, M. Mikami: 
Chem. Phys. Lett. \textbf{335}, 435--439 (2001).
\bibitem{ISNO} Y. Ishikawa, Y. Sugita, T. Nishikawa, Y. Okamoto: 
Chem. Phys. Lett. \textbf{333}, 199--206 (2001).

\bibitem{Gar} A.E. Garcia, K.Y. Sanbonmatsu: 
Proteins \textbf{42}, 345--354 (2001).

\bibitem{ZBG01} R.H. Zhou, B.J. Berne, R. Germain: 
Proc. Natl. Acad. Sci. U.S.A. \textbf{98}, 14931--14936 (2001).


\bibitem{GS02} A.E. Garcia, K.Y. Sanbonmatsu: 
Proc. Natl. Acad. Sci. U.S.A. \textbf{99}, 2782--2787 (2002).

\bibitem{ZB02} R.H. Zhou, B.J. Berne: 
{\it Proc. Natl. Acad. Sci. U.S.A.} {\bf 99}, 12777--12782 (2002).

\bibitem{FMB03} M. Feig, A.D. MacKerell, C.L. Brooks, III:
J. Phys. Chem. B \textbf{107}, 2831--2836 (2003).

\bibitem{RH03} Y.M. Rhee, V.S. Pande: 
Biophys. J. {\bf 84}, 775--786 (2003).



\bibitem{PS03} J.W. Pitera, W. Swope: 
Proc. Natl. Acad. Sci. U.S.A. \textbf{100}, 7587--7592 (2003).

\bibitem{FE03} M.K. Fenwick, F.A. Escobedo: 
Biopolymers \textbf{68}, 160--177 (2003).



\bibitem{XB00} H.F. Xu, B.J. Berne: 
J. Chem. Phys. \textbf{112}, 2701--2708 (2000). 
\bibitem{FYD02} R. Faller, Q. Yan, J.J. de Pablo: 
J. Chem. Phys. {\bf 116}, 5419--5423 (2002).

\bibitem{STREM} A. Mitsutake, Y. Okamoto: 
J. Chem. Phys. \textbf{121}, 2491--2504 (2004).

\bibitem{FE03b} M.K. Fenwick, F.A. Escobedo:
J. Chem. Phys. \textbf{119}, 11998--12010 (2003).

\bibitem{Huk2} K. Hukushima: 
Phys. Rev. E \textbf{60}, 3606--3614 (1999). 
\bibitem{WBS02} T.W. Whitfield, L. Bu, J.E. Straub: 
Physica A \textbf{305}, 157--171 (2002).
    
\bibitem{KH05} W. Kwak, U.H.E. Hansmann:
Phys. Rev. Lett. \textbf{95}, 138102 (2005).
   
\bibitem{Metro} N. Metropolis, A.W. Rosenbluth, M.N. Rosenbluth, 
A.H. Teller, E. Teller: J. Chem. Phys. \textbf{21},
1087--1092 (1953).

\bibitem{nose84}
S. Nos\'e:
Mol. Phys. \textbf{52}, 255--268 (1984).
%
\bibitem{nosejcp84}
S. Nos\'e:
J. Chem. Phys. \textbf{81}, 511--519 (1984).

 


\bibitem{BergTXT} B.A. Berg: {\it Markov Chain Monte Carlo Simulations
and Their Statistical Analysis} (World Scientific, Singapore, 2004) p. 253.

\bibitem{BergLog} B.A. Berg: 
Comp. Phys. Commun. \textbf{153}, 
397--406 (2003).
   
\bibitem{SOL1} T. Ooi, M. Oobatake, G. N\'{e}methy, H.A. Scheraga:
Proc. Natl. Acad. Sci. USA \textbf{84}, 3086--3090 (1987).

\bibitem{KONF1} H. Kawai, Y. Okamoto, M. Fukugita,
T. Nakazawa, T. Kikuchi: Chem. Lett. \textbf{1991}, 213--216 (1991).

\bibitem{KONF2} Y. Okamoto, M. Fukugita, T. Nakazawa, H. Kawai:
Protein Eng. \textbf{4}, 639--647 (1991).

\bibitem{SOL2} M. Masuya, manuscript in preparation; see also
http://biocomputing.cc/nsol/.


\bibitem{SO4} Y. Sugita, Y. Okamoto: unpublished. 

\bibitem{AMBER96} P.A. Kollman, R. Dixon, W. Cornell, T. Fox,
C. Chipot, A. Pohorille: in {\it Computer Simulation of Biomolecular Systems Vol. 3}, 
ed by A. Wilkinson, P. Weiner, W.F. van Gunsteren (Kluwer, Dordrecht, 1997) pp. 83--96.

\bibitem{TIP3P} W.L. Jorgensen, J. Chandrasekhar, J.D. Madura, R.W. Impey,
M.L. Klein: J. Chem. Phys. \textbf{79}, 926--935 (1983).

\bibitem{SK98} Y. Sugita, A. Kitao: Proteins \textbf{30}, 388--400  (1998).

\bibitem{kitao98} A. Kitao, S. Hayward, N. G\={o}: Proteins \textbf{33}, 496--517 (1998).

\bibitem{PRESTO} K. Morikami, T. Nakai, A. Kidera, M. Saito, H. Nakamura:
Comp. Chem. \textbf{16}, 243--248  (1992).
   
\bibitem{SO5} Y. Sugita, Y. Okamoto: 
Biophys. J. \textbf{88}, 3180--3190 (2005).

\bibitem{shoe87} K.R. Shoemaker, P.S. Kim, E.J. York, J.M. Stewart, R.L. Baldwin:
Nature \textbf{326}, 563--567 (1987).

\bibitem{shoe90} K.R. Shoemaker, R. Fairman, D.A. Schultz, A.D. Robertson, E.J.
York, J.M. Stewart, R.L. Baldwin: Biopolymers \textbf{29}, 1--11 (1990).



\bibitem{AMBER99} J. Wang, P. Cieplak, P.A. Kollman: 
J. Comput. Chem. \textbf{21}, 1049-1074 (2000).

\bibitem{yso04} T. Yoda, Y. Sugita, Y. Okamoto:
Chem. Phys. Lett. \textbf{386}, 460--467 (2004).


\bibitem{Molscript} P.J. Kraulis: J. Appl. Crystallogr. \textbf{24}, 946--950 (1991).

\bibitem{Raster3D} E.A. Merritt, D.J. Bacon: Methods Enzymol. \textbf{277}, 505--524 (1997).

\bibitem{yso07} T. Yoda, Y. Sugita, Y. Okamoto:
Proteins \textbf{66}, 846--859 (2007).

\bibitem{NSMO} T. Nagasima, Y. Sugita, A. Mitsutake, Y.
Okamoto: in preparation.

\bibitem{NSMO02} T. Nagasima, Y. Sugita, A. Mitsutake, Y.
Okamoto: Comp. Phys. Commun. \textbf{146},
69--76 (2002).

\bibitem{O03} Y. Okamoto: 
Metropolis algorithms in generalized ensemble.
In: {\it The Monte Carlo Method in the Physical Sciences:
Celebrating the 50th Anniversary of the Metropolis Algorithm}, 
ed by J.E. Gubernatis
(American Institute of Physics, Melville, 2003) pp. 248--260; 
e-print: cond-mat/0308119.


\end{thebibliography}
\end{document}